\documentclass[letter,doc,natbib,donotrepeattitle,floatsintext,noextraspace]{apa6}

\usepackage[english]{babel}
\usepackage[utf8x]{inputenc}
\usepackage{amsthm,amsmath,bm,bbm}
\usepackage{amssymb,mathtools}
\usepackage{graphicx}
\usepackage[colorinlistoftodos]{todonotes}
\usepackage{setspace}
\usepackage[font=singlespacing]{caption}
\usepackage{multirow}

\usepackage{etoolbox}

\usepackage{epstopdf}
\AppendGraphicsExtensions{.tif}

\usepackage{soul}
\usepackage{enumitem}

\usepackage{pdfpages}

\allowdisplaybreaks

\AtBeginDocument{%
 \abovedisplayskip=2pt plus 1pt minus 1pt
 \abovedisplayshortskip=0pt plus 1pt minus 1pt
 \belowdisplayskip=2pt plus 1pt minus 1pt
 \belowdisplayshortskip=0pt plus 1pt minus 1pt
}

\def\independent{\perp\!\!\!\perp}
\def\E{\text{E}}
\def\P{\text{P}}
\def\var{\text{var}}

\theoremstyle{plain}
\newtheorem*{theorem*}{Theorem}
\newtheorem{corollary}{Corollary}

\title{Propensity score analysis with latent covariates: Measurement error bias correction using the covariate's posterior mean, aka the \textit{inclusive} factor score}
\shorttitle{Propensity score analysis with latent covariates}
\author{\normalsize Trang Quynh Nguyen, Elizabeth A. Stuart}
\affiliation{\normalsize Johns Hopkins Bloomberg School of Public Health\\~\\\textsc{accepted by Journal of Educational and Behavioral Statistics}\\\textsc{(with supplementary material)}}
\authornote{We are indebted to the three anonymous Reviewers and the Editor for pointing us to important related work, and for their insightful comments and questions, which helped us evolve in our understanding of the problem and the proposed method, leading to substantial improvements in the content and clarity of the paper. TQN thank Daniel Scharfstein, Ilya
Shpitser and Betsy Ogburn for their reactions to this problem, which are illuminating and helpful; and thank
Abhirup Datta for indulging her struggle with probability and the $Q$ function. This work is supported by
grant R01MH099010 (PI Stuart) from the National Institute of Mental Health. Contact: trang.nguyen@jhu.edu.}

\abstract{We address measurement error bias in propensity score (PS) analysis due to covariates that are latent variables. In the setting where latent covariate $X$ is measured via multiple error-prone items $\bm W$, PS analysis using several proxies for $X$ -- the $\bm W$ items themselves, a summary score (mean/sum of the items), or the conventional factor score (cFS , i.e., predicted value of $X$ based on the measurement model) -- often results in biased estimation of the causal effect, because balancing the proxy (between exposure conditions) does not balance $X$. We propose an improved proxy: the conditional mean of $X$ given the combination of $\bm W$, the observed covariates $Z$, and exposure $A$, denoted $X_{WZA}$. The theoretical support is that balancing $X_{WZA}$ (e.g., via weighting or matching) implies balancing the mean of $X$. For a latent $X$, we estimate $X_{WZA}$ by the inclusive factor score (iFS) – predicted value of $X$ from a structural equation model that captures the joint distribution of $(X,\bm W,A)$ given $Z$. Simulation shows that PS analysis using the iFS substantially improves balance on the first five moments of $X$ and reduces bias in the estimated causal effect. Hence, within the proxy variables approach, we recommend this proxy over existing ones. We connect this proxy method to known results about weighting/matching functions \citep{McCaffrey2013,Lockwood2016}. We illustrate the method 
in handling latent covariates when estimating the effect of out-of-school suspension on risk of later police arrests using Add Health data.}


\begin{document}
\pagenumbering{gobble}
\maketitle
\pagenumbering{arabic}

\section{Introduction}

\noindent Propensity score (PS) methods are useful for estimating the effect of an exposure (or treatment) on an outcome using observational data, assuming that covariates that confound their relationship are observed. The PS -- the probability of exposure given covariates -- is a balancing score, meaning conditional on it (e.g., via matching or weighting), there is balance between exposure conditions on the covariates, thus removing confounding by them \citep{Rosenbaum1983a}.
Implicit in this no unobserved confounding assumption is the assumption that covariates are measured without error; measurement error often leads to bias in the estimated causal effect \citep{Jakubowski2015,Pearl2010,Raykov2012c,Steiner2011,Yi2012}. In PS analysis, covariate measurement error biases the estimation of the PS, resulting in residual imbalance of the true covariate (after matching or weighting), which may lead to bias in the estimated causal effect. 

We address measurement error bias when a confounder is \textit{a latent continuous variable with multiple error-prone measurements}.%
\footnote{This is not the situation where the confounder is the measurement of a latent variable (e.g., intervention decisions in education are based on the teacher's rating of the child's academic aptitude), in which case there is no measurement error bias (the rating is perfectly captured) \citep{Pohl2016}.}
(Here ``latent variable'' is a construct considered real but is not directly observable, e.g., readiness to learn.) 
This measurement error type is ubiquitous in health and education.
For example, in PS analysis evaluating the effect of retention in first grade on future academic achievement, \citet{Wu2008} incorporated covariates such as academic aptitude, personality traits, social adjustment, peer relations, and family adversity -- measured via multi-item scales.
Studying the effect of adolescent cannabis use on adulthood depression, \citet{Harder2008a} PS-matched users and non-users on measures of concentration difficulty, behavior problems, shyness, depression, anxiety, and parental supervision and monitoring, among others.

\subsection{The proxy variable approach: summary scores and factor scores}

\noindent In PS analysis practice, perhaps the most common (and simplest) way to deal with a latent covariate with multiple measurement items is to use a summary score (usually the sum or mean of the items, also called \textit{scale score} if the items are from an established scale) to represent the latent variable, and treat it as an observed covariate. Since the correlation between the summary score and the true latent variable is less than 1, there is measurement error. If the latent variable is an important confounder, such measurement error may result in appreciable bias in the estimated causal effect \citep{Steiner2011}.

For regression analysis where the interest is in estimating regression coefficients, a popular solution to this measurement error bias problem is latent variable modeling \citep{Bollen1989}: establishing a measurement model for the latent variable and then fitting a structural equation model (SEM) that combines the measurement component and the regression component with the latent and observed predictors. This does not help PS analysis, however, because although the exposure assignment model may be fit as a SEM, PS computation for each individual requires input values for all predictors, including the latent one. If we wish to use the standard procedure -- estimating the exposure assignment model (the PS model), computing the PS, then matching or weighting based on that PS -- we need a proxy for the latent variable, and desirably a better one than the summary score.

\cite{Raykov2012c} suggested using the factor score (FS) from the measurement model as a proxy for the latent variable in estimating the PS. This FS is the model predicted value of the latent variable given the measurement items. We will call it the \textit{conventional} FS (abbreviated cFS), as the term \textit{factor scores} originally referred to values generated from measurement models (called \textit{factor models}). \citeauthor{Raykov2012c} argued, intuitively, that since the cFS better represents the latent covariate, adjusting for it and the PS based on it should produce less bias than adjusting for the measurement items and the PS based on them.
Later \citet{Jakubowski2015} used simulation to compare the use of the cFS (in PS matching) to using the measurement items directly. While the author's conclusions seem to favor the cFS, our reading of the simulation results is that these two methods have similar bias.

With the benefit of hindsight, this similar bias result is not surprising, as the two methods essentially capture the same information about the latent variable. Related to this point, \citet[][pg 306]{Bollen1989} commented that the estimated FS is a weighted combination of the measurement items, and as such does not remove measurement error.
Also, from a missing data perspective, the cFS can be seen as an imputation for the latent covariate, whose imputation model relies only on the measurement items. This is a mismatch with the PS model, which relies on the exposure-confounders joint distribution; and it is well known that incompatibility of imputation and analysis models results in bias \citep{Meng1994}.

We investigate another FS that improves upon the cFS as proxy for the latent variable: the FS generated from a SEM that combines the measurement component and the exposure assignment model, thus is informed by the exposure-confounders joint distribution. We call it the \textit{inclusive} FS (abbreviated iFS), borrowing the descriptor \textit{inclusive} from \cite{Collins2001b}, who refer to missing data methods that make use of auxiliary variables as inclusive.%
\footnote{Strictly speaking, from an imputation perspective, the exposure variable and the other confounders are part of the analysis model and thus are not auxiliary. But from a conventional measurement perspective, with FSs having originated from factor analysis, these variables are auxiliary to the measurement model.}
The iFS estimates the conditional expectation of the latent variable given the measurements, the other confounders, and the assigned exposure.
We show that the iFS outperforms the cFS and the summary score in helping balance covariates in PS analysis, reducing bias in the estimated causal effect.

As the purpose of a better proxy is to improve covariate balance, its estimation belongs in the \textit{design} of the observational study, which covers ``all contemplating, collecting, organizing, and analysing of data that takes place prior to seeing any outcome data'' \citep{Rubin2007}. Hence we do not use the outcome to inform the proxy variable. This puts our method in a different class from methods that rely solely on modeling the outcome and methods that solve the joint distribution or covariance matrix of all the observed variables for a causal effect formula \citep{Pearl2010,Kuroki2014,Cai2008}.

\subsection{The weighting/matching function approach}

\noindent The proxy variable approach has an intuitive and simple appeal -- it makes sense to seek a proxy variable (a function of observed data) to stand in for just the one variable we do not observe, and once the proxy is obtained, we proceed with analysis as usual. A different approach, the weighting/matching function approach \citep{McCaffrey2013,Lockwood2016}, does not seek proxies for the unobserved variable. Instead the focus is on functions of observed (and possibly simulated) data that when used for weighting or matching result in unbiased estimation of the causal effect. While the proxy variable approach seeks a proxy for the latent variable that results in a better weighting/matching function, the weighting/matching function approach directly seeks a good weighting/matching function. Conceptually, the latter considers a larger space for weighting/matching functions, while the former places restrictions on this space, admitting only weighting/matching functions that follow from proxy variables for the latent variable. The weighting/matching function approach in principle may result in solutions that are more correct, but it is also generally more complicated.
The current paper takes the proxy variable approach, aiming to offer a method that is easy to implement for researchers who are not statisticians. We will compare the iFS proxy method to the weighting/matching function approach, focusing on special cases where closed form solutions exist for the latter.

\subsection{Data example}

\noindent Method illustration will be based on an analysis using data from the National Longitudinal Study of Adolescent to Adult Health (Add Health) \citep{Harris2013} to evaluate whether out-of-school suspension in adolescence increases the risk of subsequent problems with the law, specifically being arrested by law officers.
Since the exposure was not randomized, we use PS weighting to balance the exposed and unexposed groups on a set of baseline covariates considered potential confounders of the exposure-outcome association.
These include the latent constructs academic achievement (measured by grades for several subjects) and violence (measured by items about physical fights and weapon use).

\medskip

Next in this paper, Section 2 introduces the setting and assumptions. Section 3 provides the theoretical rationale for the proposed proxy variable. Section 4 discusses its identification and estimation. Section 5 presents simulations comparing the iFS proxy to non-inclusive proxies, with correctly specified models. Section 6 relates this method to weighting and matching functions. Section 7 evalutes the iFS proxy with misspecified models. Section 8 presents real data illustration. Section 9 closes with a discussion.

\section{Setting, Notation and Key Assumptions}

\noindent In an observational study, $A$ is a binary exposure, $Y^{(1)}$ and $Y^{(0)}$ the potential outcomes \citep{Rubin1974} under exposure and non-exposure, and $Y$ the observed outcome; $Y=AY^{(1)}+(1\!-\!A)Y^{(0)}$. The causal effect of the exposure for individual $i$ is $Y_i^{(1)}\!-Y_i^{(0)}$. Often an average effect is of interest, e.g., the average causal effect (ACE), 
$\E[Y^{(1)}\!-Y^{(0)}]$, or the average causal effect on the exposed (ACEE), 
$\E[Y^{(1)}\!-Y^{(0)}\mid A=1]$. For simplicity, we take the estimand to be the ACE, but the proposed method applies to either estimand. 

We make the usual causal inference assumptions: \textit{SUTVA} (i.e., no interference and treatment variation irrelevance) \citep{Rubin1980}; \textit{unconfoundedness} \citep{Imbens2008} given the combination of observed ($Z$) and latent ($X$) covariates (i.e., $A\independent Y^{(a)}\mid Z,X$ for $a=0,1$); and \textit{positivity} (i.e., $0<\P(A\!=\!1|Z,X)<1$). $Z$ and $X$ are often multivariate (we use univariate notation for simplicity) and may be associated due to common causes.

Had $X$ been observed, we would have been able to work directly with $Z,X$ to balance these variables between the two exposure conditions. This could be done by estimating the PS for each unit based on these variables, $e(Z,X)=\P(A=1|Z,X)$, and reweighting the units using inverse probability weights $A[e(Z,X)]^{-1}+(1\!-\!A)[1\!-\!e(Z,X)]^{-1}$ -- so that both the exposed and unexposed groups mimic the covariate distribution of the full sample.%
\footnote{If the estimand is the ACEE, odds weighting is used instead: exposed units are unweighted, and unexposed units are weighted by the odds of exposure assignment given their covariate values.}
Covariate balance may also be obtained through matching on $(Z,X)$ or on $e(Z,X)$. Once balance is obtained, the difference in mean outcome between the two exposure conditions is a valid estimate of the causal effect. Alternatively, covariate balancing may be combined with regression adjustment. Our current focus is covariate balancing, but since combination with regression adjustment is common in practice, we will comment on this where relevant.

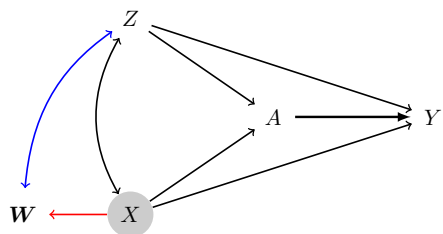
\begin{figure}[!t]
\caption{Assumed model that satisfies unconfoundedness and strong surrogacy}\label{fig:UnconfoundednessStrongSurrogacy}
\centering
\resizebox{.4\textwidth}{!}{%
\begin{tikzpicture}[
    obs/.style={circle, text centered},
    lat/.style={circle, fill=black!20, text centered},
    note/.style={rectangle, anchor=west}
]
    \node[obs] (Z) {$Z$};
    \node[lat] (X) [below=of Z, yshift=-16mm] {$X$};
    \node[obs] (A) [right=of Z, xshift=7mm, yshift=-17mm] {$A$};
    \node[obs] (Y) [right=of A, xshift=10mm] {$Y$};
    \draw[very thick, ->, >={latex}] (A) -- (Y);
    \draw[thick, ->] (Z) -- (A);
    \draw[thick, ->] (X) -- (A);
    \draw[thick, <->] (Z) to [out=-120, in=120] (X);
    \draw[thick, ->] (Z) -- (Y);
    \draw[thick, ->] (X) -- (Y);
    \node[obs] (W) [left=of X] {$\bm W$};
    \draw[thick, color=red, ->] (X) -- (W);
    \draw[thick, color=blue, <->] (Z) to [out=-145, in=85] (W);
\end{tikzpicture}%
}
\end{figure}

However, $X$ is a latent variable.  Instead of $X$, we observe K measurement items, $\bm W=(W_1,\dots,W_\text{K})$, which are functions of $X$ and noise. As we are considering FSs, which are estimates of conditional expectations of $X$ (clarified shortly), it is convenient to abbreviate conditional expectations. Let $X_W=\E[X|\bm W]$, $X_{WZA}=\E[X|\bm W,Z,A]$, etc. There is one PS version for each proxy of $X$. Using the usual PS notation  $e(\cdot)=\P(A=1|\cdot)$, the PSs estimated based on the cFS, the iFS, the summary score, and all measurement items, are estimates of $e(Z,X_W)$, $e(Z,X_{WZA})$, $e(Z,\overline W)$ and $e(Z,\bm W)$, respectively.

Some restrictions are needed on $\bm W$ to avoid contradicting the unconfoundedness assumption. Following \citet{Lockwood2016}, we assume \textit{strong surrogacy}, i.e., conditional on the covariates that satisfy unconfoundedness, $\bm W$ is independent of exposure assignment and potential outcomes, $\bm W\independent A,Y^{(a)}\mid Z,X$. Fig. \ref{fig:UnconfoundednessStrongSurrogacy} represents this setting. (In the special case where $\bm W$ are sums of $X$ and error terms, this means the error terms may depend on covariates $X,Z$, but conditional on these variables do not carry any information about exposure assignment and potential outcomes. This includes the even more special case of classical measurement error, where error terms are independent of $Z,X,A,Y^{(1)},Y^{(0)}$.) While $\bm W$ is seen primarily as measurement of $X$, we do not rule out possible association between $\bm W$ and the other covariates $Z$ conditional on $X$.

We keep with strong surrogacy in this paper for simplicity of presentation, but note that the proposed method also applies to the case where some $\bm W$ items have a direct effect on exposure assignment, where only \textit{weak surrogacy} is satisfied, i.e., $\bm W\independent Y^{(a)}\mid Z,X,A$. Application of the iFS method to this case only involves modifying the model used to generate the iFS to reflect that direct effect and requires that there is sufficient conditional independence for the model to be identified (explained later). Our approach is also relevant if $A$ influences some $\bm W$ items, which is another case of \textit{weak surrogacy}. However, this case has complicated temporal order (post-exposure measurement of pre-exposure covariate), which deserves separate consideration, so we exclude it from this paper.

Note that both strong and weak surrogacy restricts $\bm W$ to be independent of $Y^{(a)}$ given $Z,X$ or $Z,X,A$, even though this is not required to maintain unconfoundedness. This restriction is important, however, as it protects us from inducing confounding via $\bm W$ due to collider bias when matching or weighting on functions of $(\bm W,Z,A)$.

While additional assumptions will be needed for identification and estimation of the proxy variable of interest, we put off discussing them until later in the paper. Our first step is to consider the theoretical support for targeting $X_{WZA}$ as proxy for $X$.

\section{Theoretical Support  for $X_{WZA}$ as Proxy for $X$ for Covariate Balancing}

\noindent To judge whether $X_{WZA}$ is a worthwhile target, imagine that we do observe $X_{WZA}$ and see what follows.
If $X_{WZA}$ is observed, we can use PS weighting or matching to balance $(Z,X_{WZA})$. But then what happens to the unobserved residual $(X\!-\!X_{WZA})$? As shown in the theorem below (see proof in the Appendix), this residual has some nice properties that protects its mean-balance, i.e., the equality of its means between exposure conditions.

\begin{theorem*}\label{thm:1}
Let $X,Z,\bm{W}$ be random variables with finite variances. Let $A$ be a binary (0/1) random variable. Denote $\textup{E}[X|\bm{W},Z,A]$ by $X_{WZA}$.
Then
\begin{enumerate}[topsep=0pt,itemsep=0pt]
\item $\textup{E}[X\!-\!X_{WZA}\mid Z,X_{WZA},A]=0$.
\item $\textup{E}[X\!-\!X_{WZA}\mid A=1]=\textup{E}[X\!-\!X_{WZA}\mid A=0]=\textup{E}[X\!-\!X_{WZA}]=0$.
\item For any positive bounded scalar function $G=g(Z,X_{WZA},A)$,
$$\textup{E}[G(X\!-\!X_{WZA})\mid A=1]=\textup{E}[G(X\!-\!X_{WZA})\mid A=0]=0.$$
\item For any function $K=k(Z,X_{WZA})$ (possibly vector-valued), $$\textup{E}[X\!-\!X_{WZA}\mid K,A=1]=\textup{E}[X\!-\!X_{WZA}\mid K,A=0]=0.$$
\end{enumerate}
\end{theorem*}

\noindent The gist of this result is that $(X\!-\!X_{WZA})$ has mean zero conditional on any set of values for $(Z,X_{WZA},A)$. It thus has mean-balance in expectation, i.e., its expectation is zero in both the exposed and unexposed group, and this mean-balance remains after weighting by bounded functions of $(Z,X_{WZA},A)$ or after matching based on functions of $(Z,X_{WZA})$, because all the relevant conditional means of $(X\!-\!X_{WZA})$ are zero. The important implication of this result is stated in the Corollary below (see proof in the Appendix).

\begin{corollary}\label{col:1}
In the setting of Theorem, denote
$\textup{P}(A=1|Z,X_{WZA})$ by $e(Z,X_{WZA})$.
\begin{enumerate}[topsep=0pt,itemsep=0pt]
\item Assume $0\!<\!e(Z,X_{WZA})\!<\!1$. Let $Q=A/e(Z,X_{WZA})+(1\!-\!A)/[1\!-\!e(Z,X_{WZA})]$. Then
$$\textup{E}[AQX]=\textup{E}[(1-A)QX]=\textup{E}[X].$$
\item For $Z,X_{WZA}$ such that $0<e(Z,X_{WZA})<1$,
$$\textup{E}[X\mid Z,X_{WZA},A=1]=\textup{E}[X\mid Z,X_{WZA},A=0],\text{ and}$$
$$\textup{E}[X\mid e(Z,X_{WZA}),A=1]=\textup{E}[X\mid e(Z,X_{WZA}),A=0].$$
\end{enumerate}
\end{corollary}

\noindent Essentially, Corollary 1 says that weighting%
\footnote{The specific positivity assumption with respect to $(Z,X_{WZA})$ is required because positivity with respect to $(Z,X)$ does not necessarily translate to positivity for $(Z,X_{WZA})$.}
based on $e(Z,X_{WZA})$, and matching on $(Z,X_{WZA})$ or on $e(Z,X_{WZA})$ -- or on any one-to-one function of $e(Z,X_{WZA})$ -- help obtain balance on the mean of $X$ in expectation, in addition to balance on the distribution of the observed covariates $Z$. This is because such weighting/matching obtains balance on the distribution of $X_{WZA}$ while the mean-balance of $(X\!-\!X_{WZA})$ is preserved.

That balancing $(Z,X_{WZA})$ does not worsen mean-balance on $(X\!-\!X_{WZA})$ is similar to a known result: matching on the linear predictor $\beta'V$ (based on covariates $V$) of the PS does not create bias on linear combinations of $V$ that are uncorrelated with $\beta'V$ if the distributions of $V$ given $A=1$ and $A=0$ are normal \citep{Rubin1992b}, elliptical \citep{Rubin1992}, or certain mixtures of elliptical distributions \citep{Rubin2006}. Our finding for $(X\!-\!X_{WZA})$, however, does not require distributional assumptions, and only relies on the fact that $(X\!-\!X_{WZA})$ has mean zero given $(Z,X_{WZA},A)$.

Note that this result is specific to $X_{WZA}$. If we replace $e(Z,X_{WZA})$ with $e(Z,X_W)$, or $e(Z,\overline W)$, or $e(Z,\bm W)$, or $e(Z,X_{WZ})$, we no longer have this nice result for mean-balance on $X$. This shows that $X_{WZA}$ \textit{is superior as proxy for $X$ for the purpose of balancing $X$}.

This result entails unbiased causal effect estimation in a special case:

\begin{corollary}\label{col:2}
In the setting of Theorem and Corollary \ref{col:1}, let the variables assume the causal structure with the assumptions from the \textit{Setting, Notation and Key Assumptions} section. Also, assume that the outcome model is of the form
$$\textup{E}[Y^{(a)}|Z,X]=\beta_0+\beta_aa+\beta_z(Z)+\beta_{za}(Z)a+\beta_xX+\beta_{xa}Xa~~\text{for }a=0,1.$$ Then balancing the distribution of $Z$ and the mean of $X$ -- via weighting by function $Q$ or matching on $(Z,X_{WZA})$ or on $e(Z,X_{WZA})$ -- leads to unbiased estimation of the ACE.
\end{corollary}
\noindent This corollary (see proof in the Appendix) says that balancing the mean of $X$ (and the distribution of $Z$) results in unbiased ACE estimation if the outcome model is linear in $X$ in each exposure condition, and $X$ and $Z$ do not interact on the outcome. This is analogous to the regression calibration result that the conditional mean of an error-prone predictor given the other predictors helps recover a linear model's coefficients \citep[][pg. 44]{Carroll2006}. (This means, if one wishes to combine covariate balancing with regression adjustment, $X_{WZA}$ may be a reasonable choice for that purpose.)

Real world data almost surely do not belong in this special case, so balancing the mean of $X$ does not imply unbiased ACE estimation (which generally requires balance of covariate distributions, not just means). However, improved balance (relative to when using inferior proxy variables) is likely to reduce bias. We will show this in simulation studies.

\section{Identification and Estimation of $X_{WZA}$}

\noindent The challenge of working with latent variables is that the distribution of a latent variable is not nonparametrically identified. There are infinitely many candidates for variable $X$ that relate to the observed variables as depicted in Fig. \ref{fig:UnconfoundednessStrongSurrogacy}. To make progress, some assumptions are required about unobserved components of the joint distribution of $X,\bm W,Z,A$. To be clear, these assumptions essentially define the variable $X$ being considered in analysis \citep{Borsboom2003}. For example, an $X$ candidate that is assumed to influence $\bm W$ in a linear manner is a different variable from one assumed to influence $\bm W$ in a nonlinear manner. We draw selectively from assumptions common in latent variable modeling, but acknowledge this indeterminacy -- a point we revisit later.

A minor technical point before diving into the assumptions: As a latent variable does not have an inherent scale and location,
%
in latent variable modeling scale and location are provided \citep{Kline2016} usually by (1) fixing the latent variable's mean (or intercept given predictors), and either (2a) fixing its variance (or conditional variance given predictors) or (2b) fixing its slope coefficient in the model of one measurement item (aka a factor loading). The two options provide different scales for $X$, but both work, because what scale a covariate is on does not matter when the purpose is controlling for it. We use (1) and (2a), with the conventional values 0 for mean/intercept and 1 for variance.

As the purpose is to estimate $X_{WZA}$, we need to fit a model that contains enough information about the distribution of $X$ given $(\bm W,Z,A)$ that allows extracting $X_{WZA}$. The assumptions we adopt 
include
functional form and distributional assumptions for variables in the model, and conditional independence assumptions to limit the number of parameters.

\subsection{Sufficient conditional independence assumption}

\noindent The model for $(\bm W,Z,X,A)$ implied by Fig. \ref{fig:UnconfoundednessStrongSurrogacy} is unidentified because the $Z\leftrightarrow X\to\bm W$ and $Z\leftrightarrow\bm W$ paths compete to explain the association of $Z$ and $\bm W$, i.e., there are too many parameters. The simplest, and strictest, assumption to deal with this is \textit{full conditional independence}: conditional on $X$, $\bm W$ items are independent of one another and independent of $Z$. Under this assumption, the model is identified with a minimum of two $\bm W$ items.%
\footnote{This is related to the three-indicator rule of factor model identification \citep{Bollen1989}, as $(Z,A)$, by their association with $X$, acts as the third indicator. While including multiple variables, $(Z,A)$ are equivalent to one indicator only, because the $(Z,X,A)$ part of the model is saturated, with no conditional independence.}

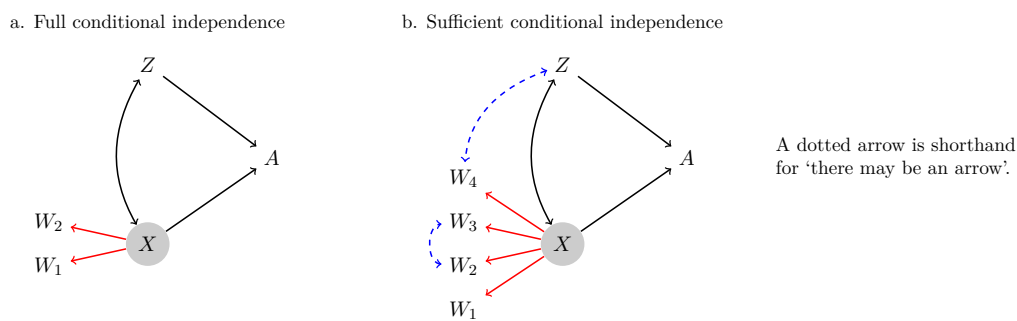
\begin{figure}[!t]
\caption{Conditional independence assumptions for identification}\label{fig:ConditionalIndependence}
\begin{center}
\resizebox{.9\textwidth}{!}{%
\begin{tikzpicture}[
    obs/.style={rectangle, text centered},
    lat/.style={circle, fill=black!20, text centered},
    anno/.style={rectangle, text centered}
]
    \node[obs] (Z) {$Z$};
    \node[lat] (X) [below=of Z, yshift=-16mm] {$X$};
    \node[obs] (A) [right=of Z, xshift=7mm, yshift=-17mm] {$A$};
    \node[anno] (t1) [above=of Z, yshift=-8mm] {\small a. Full conditional independence};
    \draw[thick, ->] (Z) -- (A);
    \draw[thick, ->] (X) -- (A);
    \draw[thick, <->] (Z) to [out=-120, in=120] (X);
    \node[obs] (Wa) [left=of X, yshift=-4mm] {$W_1$};
    \node[obs] (Wb) [left=of X, yshift=4mm] {$W_2$};
    \draw[thick, color=red, ->] (X) -- (Wa);
    \draw[thick, color=red, ->] (X) -- (Wb);
    
    \node[obs] (Z2) [right=of Z, xshift=60mm] {$Z$};
    \node[lat] (X2) [below=of Z2, yshift=-16mm] {$X$};
    \node[obs] (A2) [right=of Z2, xshift=7mm, yshift=-17mm] {$A$};
    \node[anno] (t2) [above=of Z2, yshift=-8mm] {\small b. Sufficient conditional independence};
    \draw[thick, ->] (Z2) -- (A2);
    \draw[thick, ->] (X2) -- (A2);
    \draw[thick, <->] (Z2) to [out=-120, in=120] (X2);
    \node[obs] (W2a) [left=of X2, yshift=-12mm] {$W_1$};
    \node[obs] (W2b) [left=of X2, yshift=-4mm] {$W_2$};
    \node[obs] (W2c) [left=of X2, yshift=4mm] {$W_3$};
    \node[obs] (W2d) [left=of X2, yshift=12mm] {$W_4$};
    \draw[thick, color=red, ->] (X2) -- (W2a);
    \draw[thick, color=red, ->] (X2) -- (W2b);
    \draw[thick, color=red, ->] (X2) -- (W2c);
    \draw[thick, color=red, ->] (X2) -- (W2d);
    \draw[thick, color=blue, dashed, <->] (Z2) to [out=-160, in=85] (W2d);
    \draw[thick, color=blue, dashed, <->] (W2b) [out=175, in=-175] to (W2c);
    
    \node[anno] (note) [right=of A2] {\small \begin{tabular}{l}A dotted arrow is shorthand\\for `there may be an arrow'.\end{tabular}};
\end{tikzpicture}%
}
\end{center}
\end{figure}

With more $\bm W$ items, this assumption can be relaxed, allowing for a limited number of conditional dependence relationships among $\bm W$ items, or between $\bm W$ items and $Z$. These relations, often known as \textit{residual covariances} and \textit{direct effects} in factor analysis lingo, are sometimes found through careful fitting of the measurement model \citep{Byrne2013structural} and need to be dealt with in analyses using the latent variable. Therefore, instead of full conditional independence, we only require \textit{sufficient conditional independence}, that is, there are enough conditional independence relations among $\bm W$ items, and between $\bm W$ items and $Z$, for the model to be identified. Essentially, if there are some conditional dependence relations, more $\bm W$ items are required, and conversely, the fewer $\bm W$ items there are, the fewer conditional dependence relations are allowed -- see \citet{Bollen1989} and \citet{Kline2016} on identifiability of models with latent variables. Fig. \ref{fig:ConditionalIndependence} differentiates full and sufficient conditional independence, where the dotted arrows represent possible conditional dependence.

We caution that while conditional dependence should be accommodated when it is believed to exist, it should be treated as an exception, not as the rule. While a model may be identified with just enough conditional independence and with plenty of conditional dependence, it may imply that $\bm W$ carries more information about $Z$, or about other latent variables (that account for their residual dependence) than about $X$. That would contradict the idea of $\bm W$ being measurements of $X$.

\begin{figure}
\caption{SEM used for estimation of $X_{WZA}$, with two representations of conditional dependence of measurement items}\label{fig:EstimationModel}
\begin{center}
\resizebox{.6\textwidth}{!}{%
\begin{tikzpicture}[
    obs/.style={rectangle, text centered},
    lat/.style={circle, fill=black!20, text centered},
    anno/.style={rectangle, text centered}
]
    \node[obs] (Z) {$Z$};
    \node[lat] (X) [below=of Z, yshift=-16mm] {$X$};
    \node[obs] (A) [right=of Z, xshift=7mm, yshift=-17mm] {$A$};
    \draw[thick, ->] (Z) -- (A);
    \draw[thick, ->] (X) -- (A);
    \draw[thick, ->] (Z) -- (X);
    \node[obs] (Wa) [left=of X, yshift=-12mm] {$W_1$};
    \node[obs] (Wb) [left=of X, yshift=-4mm] {$W_2$};
    \node[obs] (Wc) [left=of X, yshift=4mm] {$W_3$};
    \node[obs] (Wd) [left=of X, yshift=12mm] {$W_4$};
    \draw[thick, color=red, ->] (X) -- (Wa);
    \draw[thick, color=red, ->] (X) -- (Wb);
    \draw[thick, color=red, ->] (X) -- (Wc);
    \draw[thick, color=red, ->] (X) -- (Wd);
    \draw[thick, color=blue, dashed, ->] (Z) -- (Wd);
    \node[obs] (Ub) [left=of Wb, xshift=7mm] {\scriptsize$U_2$};
    \node[obs] (Uc) [left=of Wc, xshift=7mm] {\scriptsize$U_3$};
    \draw[thick, color=blue, ->] (Ub) -- (Wb);
    \draw[thick, color=blue, ->] (Uc) -- (Wc);
    \draw[thick, color=blue, dashed, <->] (Ub) [out=175, in=-175] to (Uc);
    \node[anno] (t1) [above=of Z, yshift=-8mm] {\small\begin{tabular}{c} b. Items conditional dependence\\as error covariances\end{tabular}};
    \node[obs] (Z2) [left=of Z, xshift=-65mm] {$Z$};
    \node[lat] (X2) [below=of Z2, yshift=-16mm] {$X$};
    \node[obs] (A2) [right=of Z2, xshift=7mm, yshift=-17mm] {$A$};
    \node[lat] (S2) [left=of X2, xshift=-16mm] {\footnotesize$S$};
    \draw[thick, ->] (Z2) -- (A2);
    \draw[thick, ->] (X2) -- (A2);
    \draw[thick, ->] (Z2) -- (X2);
    \node[obs] (W2a) [left=of X2, yshift=-12mm] {$W_1$};
    \node[obs] (W2b) [left=of X2, yshift=-4mm] {$W_2$};
    \node[obs] (W2c) [left=of X2, yshift=4mm] {$W_3$};
    \node[obs] (W2d) [left=of X2, yshift=12mm] {$W_4$};
    \draw[thick, color=red, ->] (X2) -- (W2a);
    \draw[thick, color=red, ->] (X2) -- (W2b);
    \draw[thick, color=red, ->] (X2) -- (W2c);
    \draw[thick, color=red, ->] (X2) -- (W2d);
    \draw[thick, color=blue, dashed, ->] (Z2) -- (W2d);
    \draw[thick, color=blue, dashed, ->] (S2) -- (W2c);
    \draw[thick, color=blue, dashed, ->] (S2) -- (W2b);
    \node[anno] (t2) [above=of Z2, yshift=-8mm] {\small\begin{tabular}{c}a. Items conditional dependence\\via nuisance factors\end{tabular}};
\end{tikzpicture}%
}
\end{center}
\end{figure}
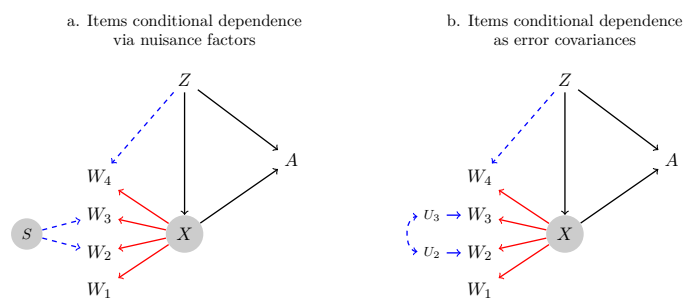

\subsection{Functional form and distributional assumptions}

\noindent We discuss these assumptions with respect to three components of the $(Z,X,A,\bm W)$ joint distribution: the covariates model, the measurement model, and the exposure assignment model. Throughout, all parameters are unknown and need to be estimated from data. Mplus \citep{Muthen2017} code is provided (in the Supplementary Material) for implementing this joint model; we use Mplus as it computes FSs based on SEMs. 

\medskip

\noindent\textbf{The covariates model.}
The assumed model so far (Fig. \ref{fig:ConditionalIndependence}b) has a curved arrow between $Z$ and $X$. To avoid having to model $Z$, which in most applications is a mix of different types of variables, we take the distribution of $Z$ as it is in the observed data, and model $X$ on $Z$. This effectively means that the SEM that is fit replaces the curved arrow with a directional arrow from $Z$ to $X$ (see Fig. \ref{fig:EstimationModel}). We are not assuming a causal effect of $Z$ on $X$, however. This is just a convenient modeling choice.

In latent variable modeling practice, a routinely made assumption is that a latent continuous variable is normally distributed, marginally or conditional on predictors. The choice to model $X$ on $Z$ means the model assumes $X$ is normal given $Z$, which allows $X$ to inherit non-normality from $Z$. This is preferable because there is usually no substantive reason to believe that $X$ is marginally normal. Also, it seems appropriate because any non-normality in $Z$ may reflect non-normality in the common causes of $Z$ and $X$ which may also influence the distribution of $X$. The covariates model is $X|Z\sim\text{N}(\alpha_zZ,~1)$. If there are more than one latent covariate (as in our application), they are treated as multivariate normal given $Z$. The SEM in Fig. \ref{fig:EstimationModel} would include, say, $X_1$ and $X_2$, each with its measurement model, and each relating to $Z$ and $A$, with a curved arrow representing their covariance given $Z$. For simplicity, we mostly refer to $X$ in singular terms.

While conditional normality is more flexible than marginal normality, it is still an arbitrary assumption, and may be undesirable in certain cases, for example, if continuous measurements demonstrate a high degree of skewness. Our simulations therefore explore cases where this assumption (along with others) is correct, and cases where it is violated.

In the Discussion section we will comment on potentials (pending investigation and development) of models that are more flexible about the distribution of $X$. For the current paper, we adopt the conditional normality assumption, because methods based on this assumption are well developed, making it easy to include conditional dependence for some measurement items if needed (see the measurement model below), and to handle multiple latent variables. It is immportant to note, though, that because $X$ is a latent variable, some assumption is needed about its distribution.%
\footnote{This setting is different from the usual setting in the measurement error literature, where $X$ is unobserved but is not a latent variable, so it may be observed elsewhere. In the latter case, if a validation dataset is available where $X$ is observed, the relationship between $X$ and a subset of the other variables (e.g., the measurement model) can be estimated in the validation sample, which may help the analysis of interest, without requiring an assumption on the distribution of $X$. For example, the \textit{conditional score} method \citep[][Ch. 7]{Carroll2006} allows fitting a generalized linear model with cannonical link when a covariate is measured with error, assuming a normal error model that is known (i.e., estimated from validation data); this method can be applied to estimate a logit exposure assignment model without assuming a distribution for $X$ \citep{McCaffrey2013}. In the current setting, on the other hand, $X$ is latent, and the measurement model and the exposure assignment model are estimated on the same data. Without imposing some structure on the distribution of $X$, this joint model is unidentified, and the measurement model alone is unidentified. A distributional assumption is the price we pay to make progress in this setting.}
\medskip

\noindent\textbf{The exposure assignment model.}
Our implementation of the method assumes a logit or probit exposure assignment model, i.e., $\P(A=1\mid Z,X)=\text{expit}(\beta_0+\beta_xX+\beta_zZ)$, or $\Phi(\beta_0+\beta_xX+\beta_zZ)$. This model assumes no interaction between the latent variable $X$ and the observed variables $Z$ on exposure assignment. If in a specific application it is believed that such an interaction plays an important role, then the current method would work less well, as it does not target covariate balance for the interaction terms.

\medskip

\noindent\textbf{The measurement model.} 
In the social and behavioral sciences, the measurements of a latent continuous variable may take a range of forms -- continuous, ordinal, count, etc. 
%
We assume that the models for $\bm W$ items given $X,Z$ are normal-linear (for continuous items) or generalized linear. In the simplest model with full conditional independence, the model for a continuous item is $W_k\mid X,Z\sim\textup{N}(\lambda_{k0}+\lambda_{kx}X,\sigma_k^2)$; that for an ordinal item with R categories is $\P(W_k> r)=\pi(-\tau_{kr}+\lambda_{kx}X)$ where $\tau_{kr}$ is the $r$th (of $\textup{R}\!-\!1$) thresholds, and $\pi(\cdot)$ is the standard normal CDF (or the expit function) if the model uses the probit (or logit) link; and the model for a count item is $\E[W_k\mid X,Z]=\exp(\lambda_{k0}+\lambda_{kx}X)$, etc. (The coefficient $\lambda_{kx}$ in these models is also referred to as a \textit{factor loading}, the \textit{loading of $W_k$ on $X$}.) Regarding the model for continuous measurements, as previously mentioned, the normal error assumption is commonly made in latent variable modeling; it is also commonly made in measurement error methods \citep[e.g.,][]{Stefanski1985,Stefanski1987}.

To accommodate conditional dependence of an item $W_k$ on $Z$, the model includes $Z$ as a predictor, so the linear predictor part of the model minus the intercept, here denoted $\theta_k$, is $\lambda_{kx}X+\lambda_{kz}Z$ instead of simply $\lambda_{kx}X$. In Fig. \ref{fig:EstimationModel}, possible conditional dependence between $\bm W$ items and $Z$ is represented by the dashed arrow from $Z$ to $\bm W$. (Again, this is a convenient modeling choice and does not represent an actual causal assumption.)

To accommodate conditional dependence between a pair or among several $\bm W$ items, we use a parameterization that attributes the source of this dependence to an unobserved common cause, represented by a nuisance latent variable independent of all other variables (see Fig. \ref{fig:EstimationModel}a). We denote this variable by $S$ (for \textit{shared} variance), and assume it is standard normal. More than one such variable may be required, for example $S_1$ to account for the conditional dependence of $W_1$ and $W_3$, and $S_2$ for the conditional dependence of $W_4,W_6$ and $W_7$. In this case, $S$ terms are added to the $\theta_k$ parts of the models for these items: $\lambda_{1s_1}S_1$ and $\lambda_{3s_1}S_1$ are respectively added to $\theta_1$ and $\theta_3$, and $\lambda_{4s_2}S_2$, $\lambda_{6s_2}S_2$ and $\lambda_{7s_2}S_2$ added to $\theta_4,\theta_6$ and $\theta_7$. (A technical detail: when only two items load on an $S$ variable, the two factor loadings are constrained to be equal, e.g., $\lambda_{1s_1}=\lambda_{3s_1}$, to pare them down to one parameter, which is appropriate as the pair represents only one dependence.) This part of the model in its most generality can be written concisely in vector/matrix form as
$$\bm\theta=\bm\lambda_xX+\bm\lambda_zZ+\bm\Lambda_s\bm S,~~\bm S\sim\textup{N}(\bm 0,\bm I),~~\bm S\independent X,Z,A,$$
where $\bm\theta$ is a vector of dimension K with each element corresponding to the model for one $\bm W$ item; $\bm\lambda_x$ and $\bm\lambda_z$ also have dimension K;  $\bm S$ is a vector of L nuisance factors, and $\bm\Lambda_s$ is a matrix of dimension K$\times$L containing the loadings of the K measurement items on these factors. In most applications, L is small if not zero, and most elements of $\bm\Lambda_s$ are zero. Likewise, most if not all elements of $\bm\lambda_z$ are zero.
This parameterization of conditional dependence among measurements is equivalent to the error covariance parameterization (see Fig. \ref{fig:EstimationModel}b)  usually used for continuous measurements -- the product of $\lambda_{1s_1}$ and $\lambda_{3s_1}$ above is the covariance of $W_1$ and $W_3$ given $X$. The nuisance $S$ parameterization, however, applies more generally to different types, and mixed types, of measurement items.

In summary, the SEM used to estimate $X_{WZA}$ (via the iFS) includes three components: (1) a linear model for $X$ given $Z$ with normal error; (2) a probit/logit model for $A$ given $Z,X$; and (3) a linear normal model, or a generalized linear model, for $\bm W$ conditional on $X,Z$, where the $\bm W$ items are for the most part independent of one another given $X,Z$, and for the most part independent of $Z$ given $X$.
In contrast, the cFS is based on a measurement model with $\bm W$ only that assumes $X$ is marginally normally distributed.

\subsection{Model fitting and FS computation}

\noindent We fit models in Mplus using maximum likelihood (ML) estimation. The iFS is then computed using the posterior mean (EAP) method \citep{Bock1981}, which is the estimated $\E[X|\bm W,Z,A]$ based on the model. Since the model may use a logit or probit link for $A$, there are two such iFS versions, which we refer to as the \textit{logit} and \textit{probit} iFS.

We also consider two approximate iFS versions. Mplus can also fit the probit model via weighted least squares, but then the iFS is computed using the posterior mode (MAP) method \citep{Muthen2004}. We refer to this iFS version as \textit{probit-WLS}, and note that it is only an approximate estimate of $\E[X|\bm W,Z,A]$. Our motivations for considering weighted least squares are practical: (1) it is computationally light, which is helpful when dealing with multiple latent variables; (2) it is Mplus's default for categorical response variables if neither estimator nor link function is specified, which may be picked up in practice as a result of users' habits. The probit-WLS iFS performs almost identically to the probit iFS in all the simulations, so we will not discuss it for the rest of the paper.

The second approximate iFS version does not require Mplus. It can be implemented with the bare minimum: software that fits linear factor models with some correlated errors and computes factor scores. This iFS is generated from a linear factor model (not a SEM) that treats $\bm W,Z,A$ all as indicators of the latent variable $X$ (ignoring the causal structure and ignoring the fact that not all these variables are continuous); replaces the effect of $Z$ on $A$ and any $Z$-$\bm W$ direct effects with error covariances; and adds error covariances between $Z$ variables. This factor model is distributionally equivalent to the SEM in Fig. \ref{fig:EstimationModel}b if that SEM uses linear models for all variables. We label this the \textit{linear} iFS.

\section{Simulation Results on the Performance of the iFS Compared to Existing Proxies for $X$ When Models Are Correctly Specified}

\noindent This section reports on simulations that confirm the theoretical result that $X_{WZA}$ is a better proxy for $X$ than the non-inclusive proxies (all $\bm W$ items, summary score, and $X_W$), and explore this proxy's relative performance in terms of balance on several moments of $X$. To zoom in on the comparison with the other proxies, we match our data generating model and estimation model so that the iFS estimates $X_{WZA}$ well and the cFS estimates $X_W$ well. A later section considers situations where the estimation model is misspecified. 

We examine the proxies' performance in PS analysis in terms of balance obtained on $Z$ and $X$, and in terms of bias, variance and mean square error (MSE) in ACE estimation. In this simulation study, we use PS weighting with inverse probability weights. That is, if $X_\text{proxy}$ is the proxy of $X$, then the estimated PS is $\hat e(Z,X_\text{proxy})=\hat\P(A=1\mid Z,X_\text{proxy})$, the corresponding weight is $\hat Q=A/\hat e(Z,X_\text{proxy})+(1-A)/[1-\hat e(Z,X_\text{proxy})]$, and the ACE is estimated by the difference between weighted mean outcomes,
$$\frac{\sum A\hat QY}{\sum A\hat Q}-\frac{\sum(1-A)\hat QY}{\sum(1-A)\hat Q}.$$

\subsection{Covariate balance}

\noindent So that the cFS model can be correctly specified, in the data generating model for this simulation, $Z$ and $X$ are multivariate normal, and $\bm W$ are independent of $Z$ given $X$. We generate (i) multivariate normal $Z,X$ with mean 0, variance 1, covariance 0, 0.4 or -0.4; (ii) two to ten continuous $\bm W$ items that are linear combinations of $X$ and noise such that their correlations with $X$ are 0.4, 0.6 or 0.8; and (iii) binary $A$ from the logit $\P(A\!=\!1|Z,X)=\text{expit}(b_0+b_1Z+b_2X)$ or probit $\P(A\!=\!1|Z,X)=\Phi(c_0+c_1Z+c_2X)$ model, where $b_1,b_2$ are 0.5 or 1, $c_1,c_2$ are 0.294 or 0.588, and $b_0,c_0$ are calibrated so that exposure prevalence ranges from 0.5 to 0.2. For each scenario, we simulate 5000 datasets of size 1000.

\begin{figure}[!t]
    \caption{Balance on the first five moments of $X$ and $Z$ given correct models. Balance shown is measured by average difference in the moment between weighted exposed and unexposed samples; and is centered at the difference obtained by analysis uzing the true $X$, which is very close to zero.}
    \centering
    \includegraphics[width=.8\textwidth]{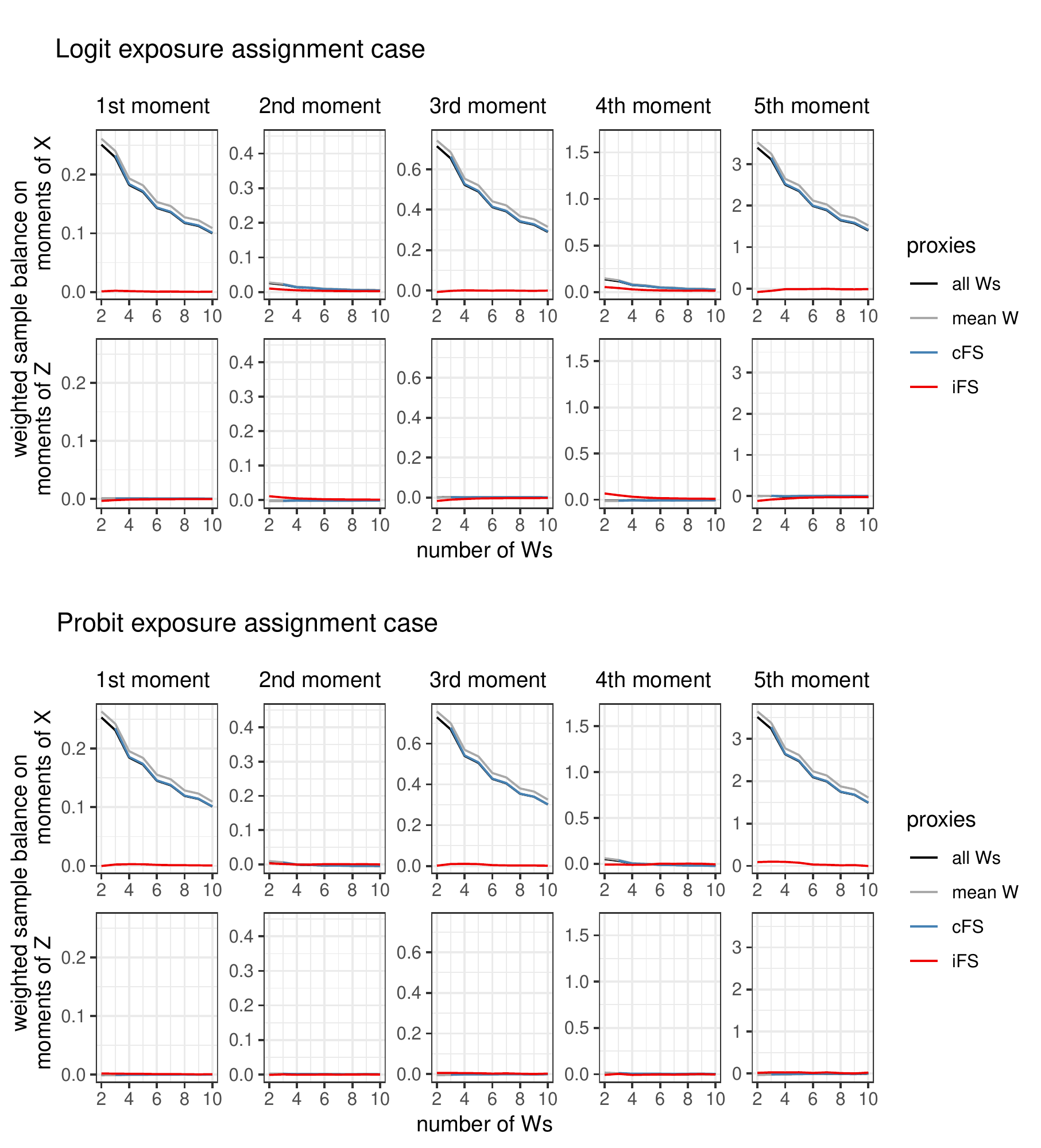}
    \label{fig:Balance}
\end{figure}

With each simulated dataset, we compute the summary score as $\overline{W}$ (the mean of $\bm W$ items) and the iFS based on the correctly specified model (logit or probit depending on the true exposure assignment mechanism). With three or more $\bm W$ items, we compute the cFS. With each of these proxies, we estimate PSs via a model with the correct link function (logit or probit based on the true model),%
\footnote{Strictly speaking, with the logit exposure assignment case, the iFS model is correctly specified, but the model form of the PS model conditional on a proxy for $X$ is only approximately correct, because logit models are not collapsible. This approximation should be close, however, as the departure of the proxy $X_{WZA}$ from $X$ is uncorrelated with both the dependent variable and the predictors in the PS model.}
and compute inverse probability weights. Other than FS estimation in Mplus, all computing is done in R \citep{R2018}. We use the R package MplusAutomation \citep{MplusAutomation2018} to bridge between R and Mplus.

Results regarding covariate balance are consistent across scenarios. We show one set of scenarios in Fig. \ref{fig:Balance} ($Z$-$X$ correlation 0.4, continuous $\bm W$, $\bm W$-$X$ correlations alternating between 0.4 and 0.6, exposure prevalence 0.3, $Z$ and $X$ having equal influence on exposure assignment with $b_1=b_2=0.5$, $c_1=c_2=0.294$).

Overall, all methods do well in achieving balance on $Z$. With respect to balance on $X$, the non-inclusive proxies do not perform well, and the iFS outperforms all of them. Essentially, \textit{the iFS obtains balance on not just the mean of $X$ but also the next four moments}. This is a better finding than was anticipated based on the theoretical result.

The imbalance in the even moments of $X$ when using the non-inclusive proxies is much less noticeable than the imbalance in the odd moments. This is due to the symmetry of the distributions of $X$ and $Z$. In fact, in scenarios where exposure prevalance is .5 or .4 (i.e., the exposure assignment model is symmetric or close to symmetric), we cannot visually detect imbalance in the even moments for any of the methods from the plot. In scenarios where exposure prevalence is smaller (e.g., .2), the imbalance in the even moments when using the non-inclusive proxies is more pronounced.

Among the non-inclusive proxies, $\overline W$ performs worse than all $\bm W$ items and worse than the cFS because here the $\bm W$-$X$ correlations are non-uniform; when $\bm W$-$X$ correlations are uniform, the non-inclusive proxy curves sit on top of one another.

\subsection{Bias reduction}

\noindent For each of the scenarios above, we simulate three outcomes. The first two are continuous, based on the linear model $Y_1\mid Z,X,A\sim\text{N}(Z\!+\!X,~4)$ and the nonlinear model $Y_2\mid Z,X,A\sim\text{N}(Z\!+\!X+\!.5X^2\!-\!.1X^3,~4)$, where the ACE is zero. The third is binary, based on the logit model $\P(Y_3=1\mid Z,X,A)=\text{expit}(A+Z+X)$,
where the ACE is $\E[\text{expit}(A+Z+X)]-\E[\text{expit}(Z+X)]$, a risk difference.

\begin{figure}[!t]
\caption{Reduction of bias in the estimated ACE when using the iFS, for three outcomes that are linear in $X$, nonlinear in $X$, and logit-linear in $X$}\label{fig:Bias}
\centering
\includegraphics[width=.6\textwidth]{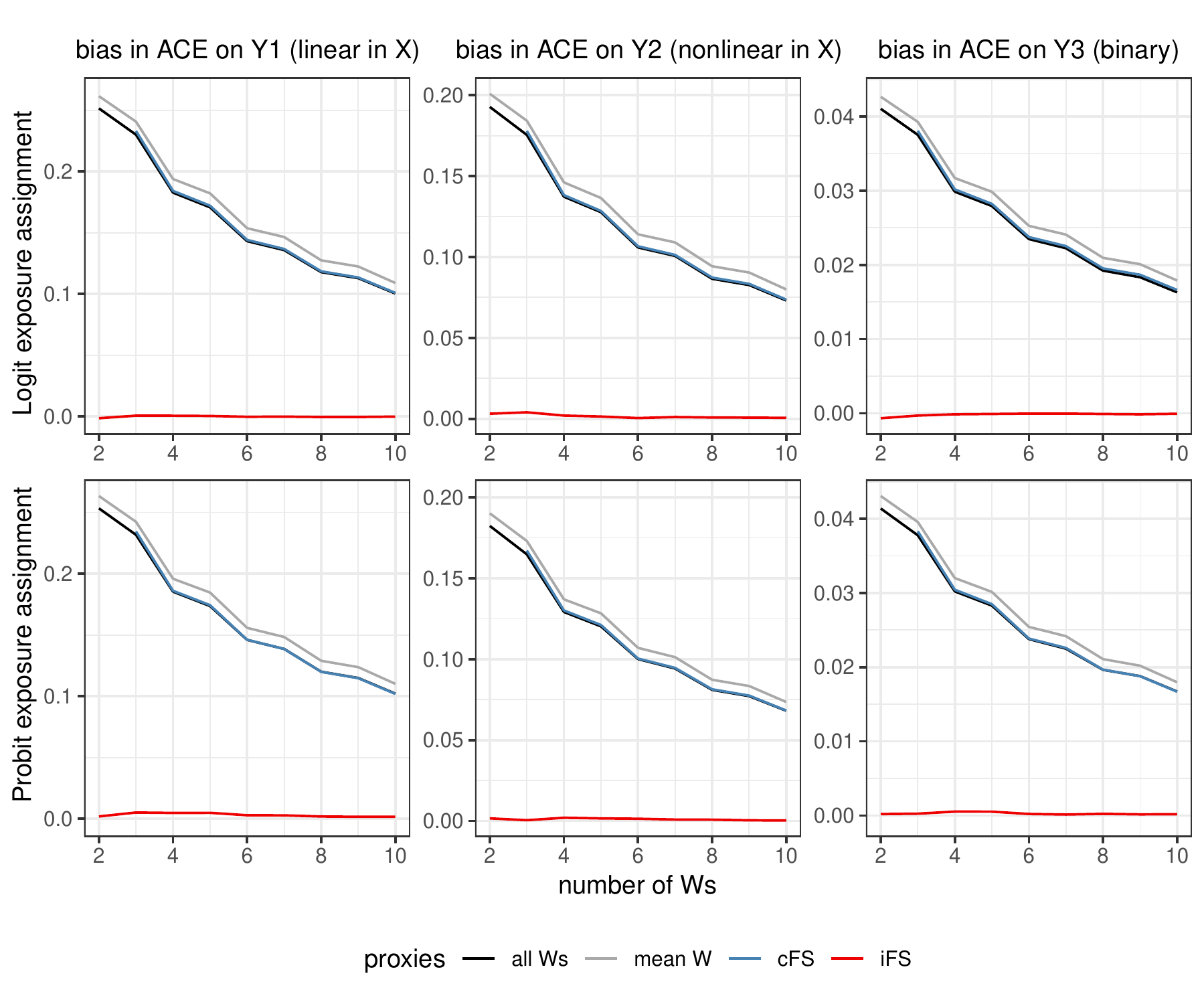}
\end{figure}

Fig. \ref{fig:Bias} presents bias in the estimated ACE. For all three outcomes bias is reduced when using the iFS as proxy for $X$. And bias seems to be reduced to zero regardless of whether the outcome is linear in $X$. This is consistent with the finding above that the iFS seems to obtain balance on several moments of $X$ and not just on the mean.

In summary, in these cases where the iFS model and the PS model are correctly specified, the iFS seems to obtain more than simply mean-balance on $X$ and bias removal in estimated ACE on outcomes linear in $X$. Simulation results suggest that the iFS achieves what looks more like distribution balance on $X$ and also bias removal in estimated ACE on outcomes nonlinear in $X$. This was not anticipated based on the theoretical result in the previous section. It is unclear whether this is only a feature of the special data generating mechanism, or whether more can be said about $X_{WZA}$ generally. We do not see a clear way to investigate this further through simulation using our current methods, because such investigation would require estimating $X_{WZA}$ well under a different data generating mechanism, which is a challenge given estimation options.

\begin{figure}[!t]
\caption{Empirical root mean square error (RMSE) and standard deviation (SD) of the ACE estimators based on the iFS and non-inclusive proxies, for the three outcomes above}\label{fig:RMSE}
\centering
\includegraphics[width=.6\textwidth]{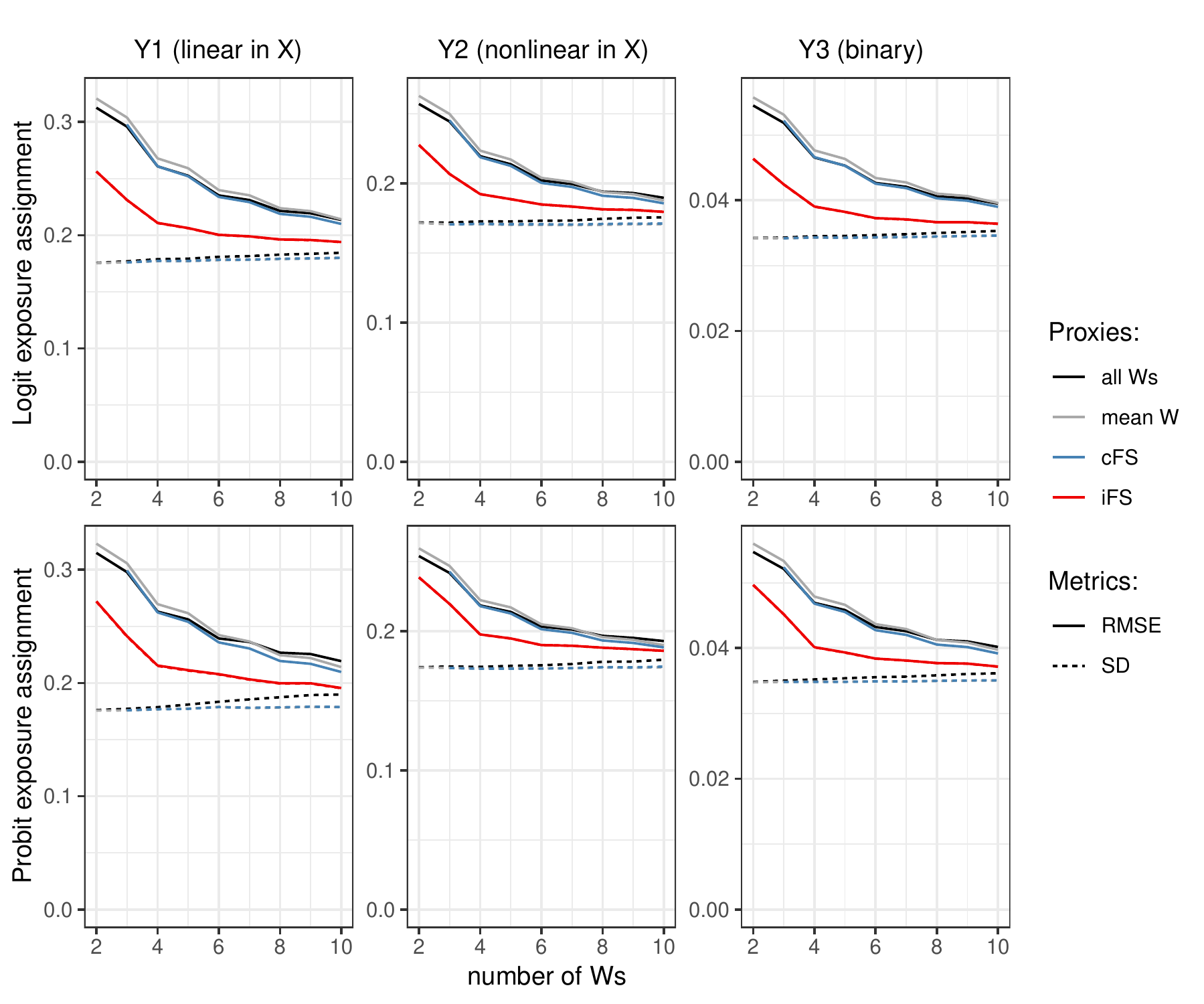}
\end{figure}

\subsection{Mean square error reduction}

\noindent Fig. \ref{fig:RMSE} shows root MSE (solid curves) and standard deviation (dashed curves) of the estimator when using the different proxies. For the iFS-based estimator, MSE reflects variance, as the red dashed curves sit on top of the red solid curves. For estimators based on non-inclusive proxies, MSE is a combination of variance and non-zero bias. While the variance of the iFS-based estimator is larger, its MSE is smaller as a result of bias removal.

\section{Connection to Known Results about Weighting and Matching Functions}

\noindent The better than expected performance of the iFS begs the question whether, given these data generating models, $X_{WZA}$ as proxy for $X$ results in a weighting/matching function that leads to unbiased ACE estimation.
This section relates this proxy to results about weighting and matching functions \citep{McCaffrey2013,Lockwood2016}.

Consider the use of weighting to adjust for $Z,X$ in estimating the ACE. The correct weighting function for this purpose is the inverse probability weight%
\footnote{The weighting function for the ACEE is $Q_0=A+(1\!-\!A)e(Z,X)/[1\!-\!e(Z,X)]$.}
based on $(Z,X)$,
$$Q_0=A[e(Z,X)]^{-1}+(1\!-\!A)[1\!-\!e(Z,X)]^{-1}.$$
With $X$ unobserved, $Q_0$ is unavailable. We consider weighting functions that are functions of observed variables $(\bm W,Z,A)$. \citet{McCaffrey2013} show that a weighting function results in unbiased ACE estimation if and only if it is unbiased for the correct weighting function. Assume such a function, denoted $Q_1$, exists. This means $\E[Q_1\mid Z,X,A]=Q_0$. 
Now consider matching instead of weighting. Correct matching functions include $H_0=(Z,X)$, or $e(Z,X)$, or any one-to-one function of $(Z,X)$ or $e(Z,X)$, none of which are available, thus we consider matching functions that are functions of observed variables. \citet{Lockwood2016} show that a matching function results in unbiased ACE estimation if and only if the weighting function based on it is unbiased for the correct weighting function. Assume that such a matching function exists. Denote it by $H_1$, and the weighting function based on it ($A[e(H_1)]^{-1}+(1\!-\!A)[1\!-\!e(H_1)]^{-1}$) by $Q_{H_1}$. This means $\E[Q_{H_1}\mid Z,X,A]=Q_0$. Weighting with a $Q_1$, or matching on a $H_1$ (if these exist and can be estimated), balances the distribution of $(Z,X)$, and obtains unbiased ACE estimation.

Our proxy variable method implies treating $H=(Z,X_{WZA})$, or $H=e(Z,X_{WZA})$, as a matching function, and treating $Q=A[e(Z,X_{WZA})]^{-1}+(1\!-\!A)[1\!-\!e(Z,X_{WZA})]^{-1}$ as a weighting function. Our theoretical result indicates that weighting with $Q$ and matching on $H$ obtains balance in the first moment of $X$. In the scenarios considered in the simulations above, weighting with $Q$ also seems to balance higher moments of $X$. Using simulation, we now examine more closely how $Q$ relates to $Q_0$ in these scenarios, and also how $Q$ and $H$ compare to $Q_1$ and $H_1$ in the scenarios with logit exposure assignment -- a special case where there are closed forms for $Q_1$ and $H_1$. We will also relate $H$ to $H_1$ in another case.

\subsection{Logit exposure assignment and normal measurement error}

\begin{figure}[!t]
    \caption{Bias of the estimated $Q_W$ (weighting function based on $\overline W$), $Q$ (weighting function based on the iFS), and $Q_1$ (unbiased weighting function), for each unit's estimated $Q_0$ (correct weight based on the true $X$)}
    \centering
    \includegraphics[width=.85\textwidth]{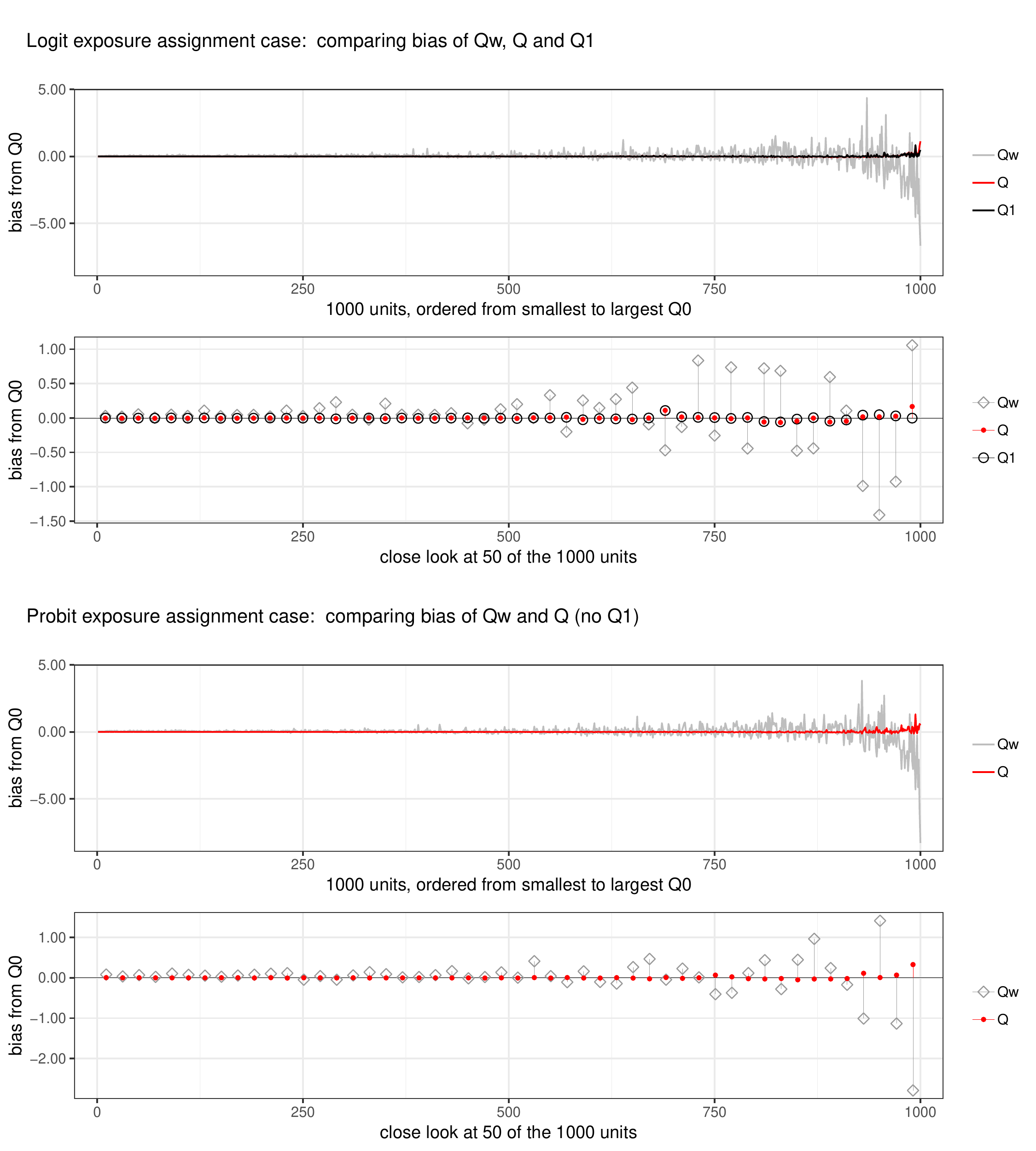}
    \label{fig:WeightBias}
\end{figure}



\noindent We start with this special case. In this case, \citet{McCaffrey2013} give a closed form for $Q_1$ for the setting with a single measurement $W$. We derive (see the Supplementary Material) an extenssion for the setting with multiple $\bm W$ items. With logit exposure assignment $\P(A=1\mid Z,X)=\text{expit}(\beta_0+\beta_zZ+\beta_xX)$, the correct weighting function is
$$Q_0=1+\exp[(1\!-\!2A)(\beta_0+\beta_xX+\beta_zZ)].$$
Given continuous measurements with normal errors $\bm W|X,Z\sim\text{N}(\bm\lambda_0+\bm\lambda_xX+\bm\lambda_zZ,~\bm\Sigma)$, the unbiased weighting function is
\begin{align*}
    Q_1=1+\exp[(1-2A)(\beta_0+\beta_xX_\text{MLE}+\beta_z Z)]\cdot\exp[-\beta_x^2\cdot\var(X-X_\text{MLE}\mid X,Z,A)/2],
\end{align*}
where $X_\text{MLE}=(\bm\lambda_x'\bm\Sigma^{-1}\!\bm\lambda_x)^{-1}\bm\lambda_x'\bm\Sigma^{-1}(\bm W-\bm\lambda_0-\bm\lambda_zZ)$ is the MLE of $X$ based on the measurement model (treating each individual's $X$ value as an unknown parameter and treating model parameters as known) and $\var(X\!-\!X_\text{MLE}|X,Z,A)=(\bm\lambda_x'\bm\Sigma^{-1}\bm\lambda_x)^{-1}$. This function is obtained by replacing $X$ in $Q_0$ with the $X_\text{MLE}$, but rescaling the exponential by an appropriate factor so that $\E[Q_1\mid Z,X,A]=Q_0$. This can be abbreviated to
$$Q_1=1+\exp[(1-2A)(\beta_0+\beta_x X^*+\beta_zZ)],$$
where $X^*=X_\text{MLE}+(2A-1)\beta_x(\bm\lambda_x'\bm\Sigma^{-1}\bm\lambda_x)^{-1}/2$, the MLE of $X$ shifted up or down a distance depending on exposure status.

We use simulation to compare $Q$ to $Q_1$ in scenarios above that involve logit exposure assignment.
With each scenario, we take one dataset of $Z,X,A$ of size 1000, and estimate for each individual $i$ their correct weight $Q_{0,i}$ via logistic regression. Conditional on the individuals' $X$ values, we generate 10,000 datasets of $\bm W$. Denote the $\bm W$ datasets by $j$, $j=1,\dots,10000$. With each dataset $j$, for each individual $i$, we estimate three weights: (i) the weight $Q_{ij}$ based on the iFS-based weighting function; (ii) a naive weight $Q_{W,ij}$ using $\overline W_{ij}$ as proxy for $X$; and (iii) the weight $Q_{1,ij}$ based on the unbiased weighting function formula (using parameter estimates from the SEM that combines the measurement and exposure assignment models). Combining the 10,000 $\bm W$ datasets, we compute, for each individual $i$, the bias (i.e., average departure from $Q_{0,i}$) and variance of each weight type.

The top half of Fig. \ref{fig:WeightBias} shows bias of the weighting functions, using one scenario%
\footnote{This scenario, which belongs in the set of scenarios represented in the top panel of the last three figures, involves three measurement items whose correlations with $X$ are 0.4, 0.6 and 0.4.}
as example; the pattern is similar across scenarios. The naive weighting function $Q_W$ (shown in gray) is biased for the correct weights. $Q_1$ (black) is unbiased as indicated by the theoretical result. Our $Q$ (red) appears unbiased for the vast majority of the units; there are only a few noticeable deviations of bias from zero for units with large correct weights. 

Notably, $Q$ mimics $Q_1$ extremely well. In the top panel, the black curve (plotted last) almost completely covers the red curve; only a tiny bit of the end of the red curve shows.%
\footnote{This might be partly a precision issue, because the iFS come from Mplus with three decimal places precision, whereas the estimated model parameters in the $Q_1$ formula have six decimal places precision.}
Also, for each unit, the variance of the $Q$ weight and that of the $Q_1$ weight are almost identical.
More interestingly, looking in specific simulated datasets, the $Q_1$ and $Q$ values are almost the same for most units, with visible differences only for units with the largest correct weights. Yet $Q$ and $Q_1$ are mathematically different functions. In $Q_1$, $\beta_0,\beta_x,\beta_z$ are coefficients of the exposure assignment model, and $X^*$ is a linear combination of $\bm W,Z,A$. In $Q=1+\exp[(1-2A)(\delta_0+\delta_1X_{WZA}+\delta_2Z)]$, $\delta_0,\delta_1,\delta_2$ are from the logit regression model regressing $A$ on $X_{WZA}$ and $Z$, and $X_{WZA}$ is a nonlinear function of $\bm W,Z,A$.

Our conclusion for this special case is that $Q$ is numerically very close to $Q_1$. Based on a result in \citet[][eq. 7.12]{Carroll2006}, $Q_1$ is an inverse probability weighting function based on $e(Z,X^*)=\P(A=1\mid Z,X^*)$, therefore $(Z,X^*)$ and $e(Z,X^*)$ are $H_1$ functions. The numerical closeness between $Q$ and $Q_1$ implies that in this special case our $H=e(Z,X_{WZA})$ is numerically very close to $H_1=e(Z,X^*)=\textup{expit}(\beta_0+\beta_xX^*+\beta_zZ)$.

\subsection{Probit exposure assignment scenarios}

\noindent With probit exposure assignment, there is no closed form for $Q_1$, but we can relate $Q$ to $Q_0$. We conduct the same simulation of $\bm W$ datasets as above (minus computation of $Q_1$ weights) for scenarios from the previous section that involve probit exposure assignment. Results are in the bottom half of Fig. \ref{fig:WeightBias}. As with the logit case, here the naive $Q_W$ is biased for $Q_0$, whereas $Q$ is unbiased for $Q_0$ for the vast majority of the units, with only noticeable non-zero bias for units whose correct weights $Q_0$ are large.

\subsection{Relating $H=(Z,X_{WZA})$ to one very special case of $H_1$}

\noindent 
There is another case with a closed form $H_1$ \citep[][Example 2]{Lockwood2016}. This case assumes that given $(X,Z)$, $\bm W$ is multivariate normal with constant covariance matrix, and that $X$ is normal given $(Z,A)$ with conditional variance not depending on $A$. Due to properties of the multivariate normal distribution, in this case $H_1$ exists and turns out to be the same as $H=(Z,X_{WZA})$, and the corresponding $Q_1$ is $Q$.
While interesting, this case is unrealistic because of the second assumption -- that $X$ is normal given $(Z,A)$. With $X$ (and $Z$) causing $A$, it must be an extremely special setting that obtains $X$ following the normal (or any other specific) distribution given $A$ (and $Z$). We do not make this assumption. (But note a related comment about the linear iFS in the next section.)

\subsection{Several observations (and speculations) based on these connections}



\noindent The weighting and matching functions implied by the proposed proxy method ($Q$ and $H$) are generally not the \textit{exact} weighting and matching functions for unbiased ACE estimation ($Q_1$ and $H_1$). 
Let us refer to $Q$ and $H$ as \textit{approximately unbiased} weighting and matching functions. (Here ``unbiased'' references unbiased estimation of the causal effect, so an unbiased weighting function is unbiased for the correct weighting function $Q_0$, but this relationship does not hold for matching functions.%
\footnote{\citet{McCaffrey2013} and \citet{Lockwood2016} label $Q_1$ and $H_1$ \textit{valid} weighting and matching functions. We opt for ``unbiased'' instead of ``valid,'' because it seems less cognitively taxing to consider something both \textit{approximately unbiased} and \textit{biased}, than both \textit{approximately valid} and \textit{invalid}.}%
)
$Q$ and $H$ are \textit{approximate} in the sense that they target balance in the first moment of $X$,%
\footnote{While simulation results show nice balance on the first five moments, based on the theoretical results, we only claim that $Q$ and $H$ target the first moment.}
while $Q_1$ and $H_1$ are \textit{exact} as they target full distributional balance (and as a result achieve unbiased effect estimation). We could think of $Q$ and $H$ as belonging in a class of approximately unbiased weighting and matching functions, in which there might be functions that are less approximate because they target balance in, say, the first two moments of $X$.  

Approximately unbiased weighting and matching functions, where available, are nice substitution for the ideal $Q_1$ and $H_1$ if these do not exist or if it is unknown whether they exist. Such approximate functions may also be relevant if $H_1$ and $Q_1$ exist but the methods for estimating them are approximate. In applications, all estimated weighting/matching functions are at best approximately unbiased in some sense; this is interesting to consider.

Whether there is pairing (or not) of a weighting function and a matching function is also interesting. Such pairing is nice to have if there is interest in estimators that, say, in the spirit of double robustness, adjust for covariates via both weighting (which requires a weighting function) and conditioning (which calls for a matching function). If a $H_1$ exists then there is a corresponding $Q_1$ \citep[the PS weight based on $H_1$ --][]{Lockwood2016}, but it is unclear whether the reverse is generally true. For example, it is not obvious what may be an $H_1$ that pairs with the $Q_1$ estimated by the procedure in section 3.1 of \citet{McCaffrey2013}. This suggests that even if a $Q_1$ is available, a search for a matching function may still be necessary. In that case, an approximately unbiased matching function, if available, is relevant as a candidate matching function. Or the form of $Q_1$ (or the procedure for estimating $Q_1$) might provide clues for potential (exact or approximate) matching functions. For example, it is the form of $Q_1$ in the logit exposure assignment case above that reveals that $(Z,X^*)$ and $e(Z,X^*)$ are unbiased matching functions. 


Our last observation is that in both the special cases with closed-form $H_1$ solutions above, the vector form $H_1$ includes $Z$ as one of the two elements, so the other element can be considered a proxy for $X$. That means the solution in these two cases belongs in the intersection of the proxy variable approach and the matching function approach. \citet{Lockwood2016} point out that a strategy for finding an unbiased weighting function $H_1$ is to find one that satisfies $A\independent(Z,X)\mid H_1$ (although this is not a necessary condition). Both $H_1=(Z,X^*)$ in the first special case and $H_1=(Z,X_{WZA})$ in the second special case satisfy this condition. That leads us to speculate that maybe in a substantial set of cases, if $H_1$ functions exist, there is a vector-valued form that contains $Z$ as an element. This points to a strategy for searching for $H_1$: searching for $X_\textup{proxy}$ such that $A\independent X\mid(Z,X_\textup{proxy})$. If such a proxy exists, then $(Z,X_\textup{proxy})$ and $e(Z,X_\textup{proxy})$ are unbiased matching functions.

\begin{figure}[!t]
    \caption{Performance of the iFS (relative to non-inclusive proxies) in cases where it does not estimate $X_{WZA}$ well. Balance (average differences in weighted sample moments) and bias measures are centered at the balance and bias levels obtained using the true $X$.}
    \centering
    \includegraphics[width=.92\textwidth]{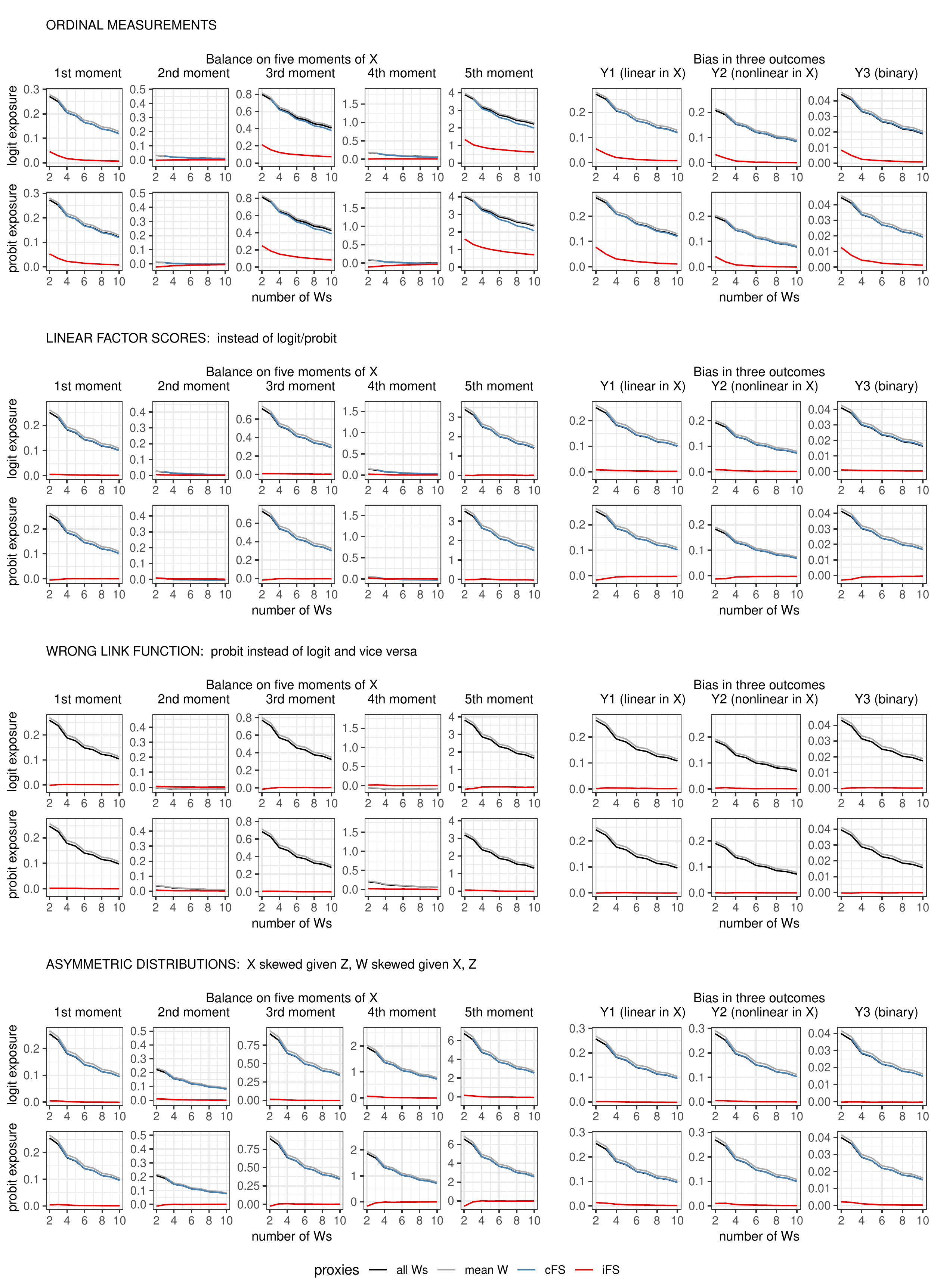}
    \label{fig:BalanceBiasIncorrect}
\end{figure}

\medskip

Before closing this section, let us tie a loose end.
Our proposal of $X_{WZA}$ as proxy for $X$ is agnostic of which estimand (e.g., ACE or ACEE) is of interest. The implied matching function $H=(Z,X_{WZA})$ does not differentiate estimands, as estimation is simply based on balancing $H$, or conditioning on and marginalizing over $H$. But so far we have discussed PS weighting only for ACE estimation. The question is, if an ACE weighting function is (approximately) unbiased for the correct ACE weighting function, whether the related ACEE weighting function is (approximately) unbiased for the correct ACEE weighting function. The answer is yes, due to a simple connection between probability and odds. For each unit $i$, the correct ACE weight, $\frac{1}{\P(A=A_i\mid Z_i,X_i)}$, is equal to 1 plus the odds of being in the other exposure condition given the unit's covariate values, $\frac{\P(A=1-A_i\mid Z_i,X_i)}{\P(A=A_i\mid Z_i,X_i)}$. That is, in the inverse probability weighted sample, each unit self-represents and in addition represents those in the other exposure condition that are similar to them. The self-representing component of the weight is always 1, and only the other-representing component varies. For an unexposed unit, the other-representing component, $\frac{\P(A=1\mid Z_i,X_i)}{\P(A=0\mid Z_i,X_i)}$, is exactly the odds weight for ACEE estimation. Therefore (approximate) unbiasedness of an ACE weighting function implies (approximate) unbiasedness of the related ACEE weighting function.

\section{Performance of the iFS Proxy When It Does Not Estimate $X_{WZA}$ Well}

\noindent In previous sections, the iFS estimates $X_{WZA}$ well. We now examine the performance of the iFS when it does not estimate $X_{WZA}$ well. Simulation results for several such cases are collected in Fig. \ref{fig:BalanceBiasIncorrect}. In all these plots, the balance (represented by average differences in weighted sample moments) and bias corresponding to each proxy are recentered by substracting values obtained when replacing the proxy with the true $X$ in the PS model.

The top panel of Fig. \ref{fig:BalanceBiasIncorrect} is the case where the iFS model is actually correctly specified, but since $\bm W$ items are ordinal with only four levels, they contain much less information about $X$ than if they were continuous. Here the iFS proxy results in residual imbalance on $X$ and bias in the ACE, but these are much reduced compared to when using non-inclusive proxies. A similar result (see the Supplemental Material) is observed in the presence of measurement errors' dependence that is not incorporated in the iFS model: this results in bias, but the bias is noticeably less than the bias when using non-inclusive proxies.

The second panel shows that the linear iFS, although misspecified, performs well as a proxy for $X$ in PS analysis. In this simulation, for all representations of $X$, the PS model uses the correct link (logit or probit). Note that this linear iFS model implies a multivariate normal structure that resembles the assumption that $X$ is normal given $(Z,A)$ in the second special case where $H_1=(Z,X_{WZA})$ in the previous section. This is interesting, but not surprising. Essentially using as approximation a suboptimal link function is equivalent to making a likely incorrect assumption. The nice result is that this approximation does not seem to matter much in these simulations.

The third panel shows the case where a wrong link function is assumed (and is used in both the FS model and the PS model), i.e., the logit link is used if the true exposure assignment mechanism is probit, and vice versa. In this case, even using the true covariate $X$ leads to some bias due to model misspecification. The plots show balance and bias for different proxies that are centered at what is achieved using the true $X$ with the wrong link function. The iFS performs very well relative to this appropriate benchmark.

The bottom panel represents the case where the true model of $X$ given $Z$ is skewed and the true model of $\bm W$ given $X$ is also skewed. The FS model thus misspecifies both of these elements. Again, the iFS outperforms the non-inclusive proxies in this setting.

\section{The Real Data Example}

\noindent We now illustrate bias correction using the iFS method with real data. The study question is whether out-of-school suspension (hereafter, suspension) in adolescence increases the risk of subsequent problems with the law. The Add Health sample is a representative sample of adolescents in the United States recruited during the 1994-95 school year (wave 1) when in grades 7-12 and followed into adulthood \citep{Harris2013}. We use data from male participants in the public access dataset for whom data are available on analysis variables. The analysis is for illustrative purposes only; results should not be taken as substantive findings.

The exposure is suspension during the approximately one year between waves 1 and 2. The outcome is police arrest after wave 2, or more precisely, between wave 2 and the last wave with data (wave 4 in 2008). Analysis is restricted to male participants who at wave 1 had prior suspension history but no prior arrests. This restriction makes exposed and unexposed individuals more similar in their chance of receiving the exposure, thus more reasonable to compare. It excludes individuals who had little to no chance of being suspended. The goal is to estimate the effect of exposure on the exposed (ACEE).

In this sample (n = 417), 140 (33.6\%) reported having the exposure. This is a small group (relative to the full original sample of several thousand), and we are not sure whether the subsetting of the data that led to this sample retains the representativeness for the corresponding subset of the national population. We thus choose to estimate the effect of the exposure on this specific group of exposed individuals, and ignore their survey weights from Add Health. Our analysis incorporates clustering (within schools) information to accommodate within-cluster correlation, using the R package survey \citep{Lumley2004surveypackage,Lumley2019surveypackage}.

Targeting the ACEE, we want to use PS weighting to weight the unexposed group to mimic the exposed group with respect to baseline covariates (measured at wave 1). These include observed covariates age, race, ethnicity, parent education, parent marital status; and latent covariates academic achievement (measured by four grades for math, English, social and natural sciences) and violence tendency (measured by four items reporting past 12-month physical fights including weapon use). The latent variables' measurement items are ordinal: academic achievement items are coded 1=grade D/F, 2=C, 3=B and 4=A; violence items are on a 0=never to 3=five-or-more-times response scale \citep{Harris2009}.

Before further analysis, one task is required to bridge from the methodological investigation thus far to analysis in practice. As the focus of the current paper is covariate balancing through PS weighting using the iFS, we have taken for granted that the causal effect is estimated simply by taking the difference in PS weighted mean outcome between exposure conditions. All our simulation to this point uses this simple method, which we refer to as the \textit{weighting-only} estimator. While this is fine for simulation studies -- as we only look at method performance over many simulated datasets but not at any single dataset specifically -- it might not work well for actual data analysis where we have only one sample and covariate balance obtained on the sample is often not exact, in which case we would worry about bias due to residual imbalance. To address this issue, we also use a second estimator, labeled \textit{weighting-plus}, which combines PS weighting with regression adjustment. The latter is done by fitting, to the weighted sample, a working outcome model which we do not assume to be correct (here a logistic model regressing outcome on exposure and covariates), computing model-predicted potential outcome probabilities, and averaging them over the inference population. In this case of estimating the ACEE for a binary outcome, the weighting-only estimator is the difference between the outcome proportion in the exposed and the odds-weighted outcome proportion in the unexposed,
$$\widehat{\text{ACEE}}_\text{weighting-only}=\frac{\sum_{i=1}^n A_iY_i}{\sum_{i=1}^n A_i}-\frac{\sum_{i=1}^n (1-A_i)\hat Q_iY_i}{\sum_{i=1}^n (1-A_i)\hat Q_i}.$$
Here $\hat Q_i$, for an unexposed individual, is the odds of exposure assignment given the individual's covariates, estimated based on the PS model. The weighting-plus estimator is equivalent to replacing the second term with the mean, taken over the exposed individuals, of their predicted potential outcome probabilities under non-exposure (denoted $\hat Y_i^{(0)}$) based on the working outcome model fitted to the odds weighted sample.
$$\widehat{\text{ACEE}}_\text{weighting-plus}=\frac{\sum_{i=1}^n A_iY_i}{\sum_{i=1}^n A_i}-\frac{\sum_{i=1}^n A_i\,\hat Y_i^{(0)}}{\sum_{i=1}^n A_i}.$$
The weighting-plus estimator is related to the \textit{standardized} estimator, which has been shown to be consistent in the randomized trial setting (without measurement error), not assuming the working outcome model is correct \citep{Rosenblum2010, Steingrimsson2017}. It is relevant to our current setting because the purpose of PS weighting is to mimic a randomized trial, and the residual imbalance we have in a PS analysis is analogous to the chance imbalance in a randomized trial. An additional simulation study for the current setting mimicking the sample data shows that the weighting-plus estimator performs well even when the working outcome model is misspecified (see the Supplemental Material). We now proceed with the data example.

Prior to PS analysis, we conduct factor analysis of the measurement items of the latent variables. Factor analysis supports unidimensionality for each set of items.
The violence set has good internal consistency, ordinal alpha \citep{Zumbo2007} = 0.81; the academic achievement set is less internally consistent, ordinal alpha = 0.67. We also conduct multi-group factor analysis to check for measurement non-invariance of the two latent variables between the exposed and unexposed groups, and conclude that measurement invariance is supported. Note that with temporal ordering (baseline covariates measured prior to the exposure), exposure status does not affect measurement. Measurement invariance could have been caused, however, by factors that precede measurement that influence both measurement and exposure assignment, and would have complicated the analysis. Fortunately measurement invariance is supported by the data.

The first three numeric columns in Table \ref{tab:balance} summarize the baseline covariates in the exposed and unexposed groups. The standardized mean differences (SMD) for most covariates have absolute values above 0.1 indicating the group means differ by more than 0.1 standard deviation. Exposed participants were more likely to be African-American; their parents on average had lower education and were less likely to be married and more likely to be single parents. Also, the mean scores (i.e., the averages of measurement items) of the latent variables are distributed differently between the two groups: those in the exposed group on average had higher violence mean scores and lower academic achievement mean scores. The same pattern is observed with the iFSs for these two latent variables.

Just for illustration, we first conduct PS weighting using the observed covariates and the \textit{mean scores} of the latent covariates, as if we did not know the iFS method. This results in improved covariate balance shown in the next two columns of Table \ref{tab:balance}: all but one of the variables put in the PS model have weighted SMD with absolute value below 0.1, indicating excellent balance \citep{Stuart2010}. A side effect is improved balance on the two iFSs, which were not included in the PS model. However, the iFS for academic achievement still retains a large SMD, which we would not know without computing the iFSs.

With the weights based on the mean scores (ignoring measurement error), we would report either of the effect estimates in the ``neither corrected'' row in Table \ref{tab:results}, and conclude that for male participants with prior history of suspension, additional suspension (between waves 1 and 2) increased the risk of subsequent arrests by police, by 11.0 percentage points (95\% CI = (3.1, 18.3)) if using the weighting-only method, or by 11.6 percentage points (95\% CI = (3.7, 18.5)) if using the weighting-plus method. The confidence intervals are equal-tail intervals obtained via the nonparametric bootstrap.

\begin{table}
\caption{\label{tab:balance}Balance of baseline covariates between exposed (n=140) and unexposed (n=277) groups (1) before PS weighting, and after PS weighting based on the latent covariates' (2) mean scores and (3) iFSs. SMD = standardized mean difference.}
\resizebox{1\textwidth}{!}{
\begin{tabular}{lrrrrrrrrrr}
& Exposed && \multicolumn{2}{c}{Unexposed group} && \multicolumn{5}{c}{Unexposed group after PS weighting}
\\ \cline{7-11}
& group && \multicolumn{2}{c}{before PS weighting} && \multicolumn{2}{c}{based on mean scores} && \multicolumn{2}{c}{and based on iFSs}
\\\cline{2-2}\cline{4-5}\cline{7-8}\cline{10-11}
& mean (\%) && ~~~mean (\%) & SMD && ~~~mean (\%) & SMD && mean (\%) & SMD
\\\hline
\textbf{Observed covariates}
\\
Age & 15.9&& 16.1 & \textcolor{red}{-0.16} && 15.9 & 0.02 && 15.8 & 0.08
\\
Race (\%)
\\
~~White & 62.9&& 64.6 & -0.04 && 64.3 & -0.03 && 66.7 & -0.08
\\
~~Black/African-American & 33.6 && 27.8 & \textcolor{red}{0.13} && 32.6 & 0.02 && 30.2 & 0.07
\\
~~Native American & 7.9 && 4.7 & \textcolor{red}{0.14} && 8.7 & -0.03 && 8.4 & -0.02
\\
~~Asian & 2.1 && 3.6 & -0.08 && 1.4 & 0.05 && 1.1 & 0.08
\\
Hispanic ethnicity (\%) & 8.6 && 10.8 & -0.08 && 12.0 & \textcolor{red}{-0.11} && 11.5 & -0.10
\\
Parent education (\%)
\\
~~Less than high school & 18.6 && 17.0 & 0.04 && 21.2 & -0.06 && 22.8 & \textcolor{red}{-0.10}
\\
~~High school & 38.6 && 26.4 & \textcolor{red}{0.27} && 35.8 & 0.06 && 35.9 &  0.05
\\
~~Business/vocational training & 15.0 && 11.9 & 0.09 && 13.4 & 0.05 && 13.4 & 0.05
\\
~~Some college (not graduated) & 12.1 && 25.6 & \textcolor{red}{-0.33} && 11.4 & 0.02 && 10.4 & 0.06
\\
~~College graduate or higher & 15.7 && 19.1 & -0.09 && 18.2 & -0.06 && 17.5 & -0.05
\\
Parent marital status (\%)
\\
~~Married & 59.3 && 66.1 & \textcolor{red}{-0.14} && 56.7 & 0.05 && 56.8 & 0.05
\\
~~Single & 15.0 && 4.3 & \textcolor{red}{0.40} && 16.5 & -0.04 && 15.7 & -0.02
\\
~~Widowed & 3.6 && 4.3 & -0.04 && 3.2 & 0.02 && 2.8 & 0.05
\\
~~Divorced & 15.0 && 20.6 & \textcolor{red}{-0.14} && 16.0 & -0.03 && 16.9 & -0.05
\\
~~Separated & 7.1 && 4.7 & \textcolor{red}{0.11} && 7.6 & -0.02 && 7.9 & -0.03
\\\hline
\textbf{Proxies of latent covariates}
\\
Violence
\\
~~Mean score (range 0-3) & 0.65 && 0.43 & \textcolor{violet}{0.39} && 0.66 & -0.02 && 0.70 & -0.09
\\
~~Inclusive factor score & -0.86 && -1.32 & \textcolor{purple}{0.55} && -0.94 & 0.08  && -0.84 & -0.03
\\
Academic achievement
\\
~~Mean score (range 1-4) & 1.18 && 1.52 & \textcolor{violet}{-0.50} && 1.17 & 0.01 && 1.02 & \textcolor{violet}{0.23}
\\
~~Inclusive factor score & -0.21&& 0.40 & \textcolor{purple}{-0.75} && 0.01 & \textcolor{purple}{-0.28} && -0.20 & -0.01
\\\hline
\end{tabular}
}
\end{table}

\begin{table}[h!]
\caption{\label{tab:results}Changes in causal estimates as a result of measurement error correction (using the iFS method) for latent covariates. Proportions and effects are on percent scale.}
\resizebox{\textwidth}{!}{
\begin{tabular}{lcccrlccrl}
&&& \multicolumn{3}{c}{WEIGHTING-ONLY ESTIMATOR} && \multicolumn{3}{c}{WEIGHTING-PLUS ESTIMATOR}
\\ \cline{4-6} \cline{8-10}
& outcome && weighted &&&& mean predicted
\\
& proportion && outcome &  &&& potential outcome & 
\\
& in the && proportion & ACEE & 95\% && probability under & ACEE & 95\%
\\
& exposed && in the & point & confidence && non-exposure & point & confidence
\\ 
& group && unexposed & estimate & interval && for the exposed & estimate & interval
\\\hline
neither corrected & 70.7 && 59.7 & 11.0 & (3.1, 18.3) && 59.1 & 11.6 & (3.7, 18.5)
\\
violence corrected & 70.7 && 61.1 & 9.7  &&& 60.5 & 10.3 
\\
acad. achiev. corrected & 70.7 && 62.5 & 8.2 &&& 61.4 & 9.3 
\\
both corrected & 70.7 &&  63.2 & 7.5 & (-0.8, 15.9) && 62.2 & 8.6 & (1.7, 18.2)
\\\hline
\end{tabular}
}
\end{table}

We now use the iFS method to correct bias due to measurement error. PS weighting based on the observed covariates and the iFSs of the latent covariates results in balance shown in the last two columns of the Table \ref{tab:balance}. Of the covariates entered in the PS model, only one has a weighted SMD with absolute value above 0.1, and only slightly so.

Note that balance in the mean score does not generally imply balance in the iFS, or vice versa. This is due to the fact that individuals with the same mean score value but different exposure status tend to have different true value on the latent covariate, and the iFS reflects this difference because it incorporates information about exposure status. This means when the iFS is balanced -- which we aim to achieve -- the mean score generally is not; the difference in mean score here depends on measurement reliability and on the strength of the latent variable's association with exposure assignment.

Using the iFSs as proxies for the latent covariates, we arrive at the result in the ``both corrected'' row in Table \ref{tab:results}. The average causal effect of additional suspension on the exposed is estimated to be an increase of the risk of subsequent police arrest of 7.5 percentage points (95\% CI = (-0.8, 15.9)) if using the weighting-only estimator, or 8.6 percentage points (95\% CI = (1.7, 18.2)) if using the weighting-plus estimator. These final effect estimates are smaller than those that do not benefit from the iFS method.

To fully illustrate how measurement error bias correction changes the estimated causal effect, two additional rows in Table \ref{tab:results} show results from analyses that uses the iFS method to correct measurement error for only one of the two latent variables (using the iFS for one but the mean score for the other one). Compared to no correction, measurement error correction results in reduction of the estimated causal effect. Correction for academic achievement only (the latent covariate with less reliable measurement) results in greater reduction in the estimated causal effect than correction for violence. Correction for both latent variables combined results in the largest reduction in the estimated causal effect.

\section{Discussion}

\noindent Our goal was to find a proxy for a latent covariate $X$ that would help reduce measurement error bias in PS analysis. The proxy we propose is the posterior mean $X_{WZA}$ of $X$ (given measurements, observed covariates and assigned exposure), estimated by the iFS via SEM. This proxy targets balance on the first moment of $X$, an improvement over non-inclusive proxies that are informed only by the measurements. In simulation, this proxy substantially improves covariate balance and reduces bias, both when model assumptions are correct and when they are violated. This is an important result given that latent variables are commonly encountered. In addition, the theoretical results for the proxy $X_{WZA}$ apply in the general measurement error setting, and are not specific to latent variables.

An interesting connection exists between our proxy variable method and the weighting and matching functions approach. \citet{McCaffrey2013} and \citet{Lockwood2016} define and search for \textit{valid} weighting and matching functions, which we refer to here as \textit{unbiased} weighting and matching functions; these target distributional balance of the latent covariate. The weighting and matching functions implied by our proxy variable, which targets balance on the first moment of the latent covariate, belongs in a class of \textit{approximately unbiased} weighting and matching functions. It would be interesting to explore, both generally and in special cases, the possibilities of other functions on a spectrum between these approximately unbiased and the exactly unbiased functions.



Turning our gaze back to the $X_{WZA}$ proxy strategy, this  study is a first step; there are potential extensions as well as gaps. One clear direction is extending the range of models that can be used that allow computing the iFS. A potential avenue is to seek ways to incorporate measurement models that are more flexible about the distribution of the latent variable, e.g., using a normal mixture or a skew-$t$ distribution \citep[which are now available for continuous measurement items -- see][]{Asparouhov2015,Wall2012,Lin2015,McLachlan2007}. Another possibility is to take the approach in \citet{Rabe-Hesketh2003} of approximating the unknown distribution of the latent variable with a discrete distribution whose number of levels, values, and probabilities are estimated to maximize the likelihood; the same approach was used by \citet{McCaffreyPoster2015}. With either approach, much needs to be done to develop tools for estimating the latent variable's posterior mean, and to allow flexibility in handling multiple latent variables and measurement items of diverse types.

Another area for future work is variance estimation, which is complicated due to the combination of $X_{WZA}$ estimation, PS estimation, and also the variance of $X\!-\!X_{WZA}$. Even the bootstrap, which we use in the data example, should be systematically investigated for the current setting.
Thinking outside the box, a possibility to be investigated that may help correct for bias while capturing full variability is Bayesian analysis where the latent variable is considered a fully missing variable and samples from its posterior are used in PS analysis; this approach does not require FS computation and thus may be more flexible.

In conclusion, obtaining valid causal effect estimates in the presence of latent confounders requires the use of methods that account for measurement error in those variables. As shown here, an easily implementable solution, the iFS method, can help reduce bias and lead to improved causal inferences. We comment on the practical application of this method and provide detailed implementation instructions and code in the Supplementary Material.

\bibliography{refs}

\appendix

\singlespacing

\setlength{\abovedisplayskip}{3pt plus 1pt minus 1pt}
\setlength{\belowdisplayskip}{3pt plus 1pt minus 1pt}
\setlength{\abovedisplayshortskip}{3pt plus 1pt minus 1pt}
\setlength{\belowdisplayshortskip}{3pt plus 1pt minus 1pt}

\section{}
\begin{proof}[Proof of Theorem]\hfill
\begin{enumerate}[itemsep=0ex]
\item[(1)] Note that $\E[X\!-\!X_{WZA}\mid\bm{W},Z,A]=0$ and $X_{WZA}$ is a function of $(\bm{W},Z,A)$.
\begin{align*}
    \E[X\!-\!X_{WZA}\mid Z,X_{WZA},A]
    &=\E\{\,\E[X\!-\!X_{WZA}\mid\bm W,Z,X_{WZA},A]\mid Z,X_{WZA},A\,\}\\
    &=\E\{\,\underbrace{\E[X\!-\!X_{WZA}\mid\bm W,Z,A]}_{0}\mid Z,X_{WZA},A\,\}
    =0.
\end{align*}
\item[(2)] 
\begin{align*}
    &\E[X\!-\!X_{WZA}\mid A]=\E\{\,\underbrace{\E[X\!-\!X_{WZA}\mid Z,X_{WZA},A]}_{=0\text{~by (1)}}\mid A\,\}
    =0.
    \\
    &\E[X\!-\!X_{WZA}]=\E\{\underbrace{\E[X\!-\!X_{WZA}\mid A]}_{=0\text{ by above line}}\}=0.
\end{align*}
\item[(3)]
\begin{align*}
    \E[g(Z,X_{WZA},A)(X\!-\!X_{WZA})\mid A]
    &=\E\{\,\E[g(Z,X_{WZA},A)(X\!-\!X_{WZA})\mid Z,X_{WZA}, A]\mid A\,\}\\
    &=\E\{\,g(Z,X_{WZA},A)\underbrace{\E[X\!-\!X_{WZA}\mid Z,X_{WZA}, A]}_{=0\text{~by (1)}}\mid A\,\}\\
    &=0~~\text{for bounded }g(Z,X_{WZA},A).
\end{align*}
\item[(4)]
\begin{align*}
    \E[X\!-\!X_{WZA}\mid k&(Z,X_{WZA}),A]\\
        &=\E\{\,\E[X\!-\!X_{WZA}\mid Z,X_{WZA},k(Z,X_{WZA}),A]\mid k(Z,X_{WZA}),A\,\}\\
        &=\E\{\,\underbrace{\E[X\!-\!X_{WZA}\mid Z,X_{WZA},A]}_{=0\text{~by (1)}}\mid k(Z,X_{WZA}),A\,\}
        =0.
\end{align*}
\end{enumerate}
\end{proof}

\begin{proof}[Proof of Corollary 1]\hfill
\begin{enumerate}
\item[(1)] We show that $\E[AQX]=\E[X]$.
\begin{align*}
    \E[AQX]
    &=\E[AQX_{WZA}]+\E[AQ(X\!-\!X_{WZA})]
    \\
    &=\E\left[\frac{A}{\P(A=1|X_{WZA})}X_{WZA}\right]+\underbrace{\E[Q(X\!-\!X_{WZA})\mid A=1]}_{=0\text{ by part (3) of Theorem}}\P(A=1)\\
    &=\E\left\{\E\left[\frac{A}{\P(A=1|X_{WZA})}X_{WZA}\mid X_{WZA}\right]\right\}
    \\
    &=\E\left[\frac{\E[A|X_{WZA}]}{\P(A=1|X_{WZA})}X_{WZA}\right]
    \\
    &=\E[X_{WZA}]
    \\
    &=\E[X]-\underbrace{\E[X\!-\!X_{WZA}]}_{=0\text{ by part (2) of Theorem}}=\E[X].
\end{align*}
\item[(2)] 
\begin{align*}
    \E[X\mid Z,X_{WZA},A]&=\E[X_{WZA}\mid Z,X_{WZA},A]+\underbrace{\E[X\!-\!X_{WZA}\mid Z,X_{WZA},A]}_{=0\text{ by part (1) of Theorem}}=X_{WZA},
    \\
    \E[X\!\mid\! e(Z,X_{WZA}),A]&=\E[X_{WZA}\mid e(Z,X_{WZA}),A]+\underbrace{\E[X\!-\!X_{WZA}\mid e(Z,X_{WZA}),A]}_{=0\text{ by part (4) of Theorem}}\\
    &=\E[X_{WZA}\mid e(Z,X_{WZA})],
\end{align*}
where the last equality is the PS's balancing property \citep{Rosenbaum1983a}.
\end{enumerate}
\end{proof}

\begin{proof}[Proof of Corollary 2]\hfill

\noindent With this specific outcome model, the ACE is $\beta_a+\E[\beta_{za}(Z)]+\beta_{xa}\E[X]$. We need to show that the weighting and matching estimators in this Corollary are unbiased for this ACE.
Denote $Y^{(a)}\!-\!\E[Y^{(a)}|Z,X]$ by $\epsilon^{(a)}$. The outcome model assumption implies $\E[\epsilon^{(a)}|Z,X]=0.$ Because $Y^{(a)}$ is a function of $(Z,X,\epsilon^{(a)})$, it follows from the unconfoundedness assumption ($A\independent Y^{(a)}\mid Z,X$) that $A\independent\epsilon^{(a)}\mid Z,X$, which implies $$\E[\epsilon^{(a)}\mid Z,X,A]=\E[\epsilon^{(a)}\mid Z,X]=0.$$
We assume $0<e(Z,X_{WZA})<1$.

\begin{enumerate}
    \item[(1)] Let's first prove the result for weighting using the weight function $Q$.
    
    Since $Q$ is a function of $(\bm W,Z,A)$, it followes from the weak surrogacy assumption ($\bm W\independent Y^{(a)}\mid Z,X,A$) that $Q\independent Y^{(a)}\mid Z,X,A$. Given the outcome model assumption, $Y^{(a)}$ is a function of $(Z,X,\epsilon^{(a)})$, it follows that $$Q\independent\epsilon^{(a)}\mid Z,X,A.$$
    
    Now consider the weighted mean outcome in each exposure arm:
    \begin{align*}
        &\E[QY\!\mid\! A=a]=\E[QY^{(a)}\mid A=a]
        \\
        &=\E\{Q[\beta_0+\beta_aa+\beta_z(Z)+\beta_{za}(Z)a+\beta_xX+\beta_{xz}Xa+\epsilon^{(a)}]\mid A=a\}
        \\
        &=\beta_0+\beta_aa+
        \\
        &~~~~\underbrace{\E[Q\beta_z(Z)|A\!=\!a]}_{\E[\beta_z(Z)]}+\underbrace{\E[Q\beta_{za}(Z)|A\!=\!a]}_{\E[\beta_{za}(Z)]}a+(\beta_x+\beta_{xa}a)\underbrace{\E[QX|A\!=\!a]}_{\E[X]\text{ by Corollary 1(1)}}+\E[Q\epsilon^{(a)}|A\!=\!a]
        \\
        &=\beta_0+\beta_aa+\E[\beta_z(Z)]+\E[\beta_{za}(Z)]a+(\beta_x+\beta_{xa}a)\E[X]+\E\{\E[Q\epsilon^{(a)}|Z,X,A\!=\!a]\mid A\!=\!a\}
        \\
        &=\beta_0+\beta_aa+\E[\beta_z(Z)]+\E[\beta_{za}(Z)]a+(\beta_x+\beta_{xa}a)\E[X]+
        \\
        &~~~~\E\{\E[Q|Z,X,A\!=\!a]\cdot\underbrace{\E[\epsilon^{(a)}|Z,X,A\!=\!a]}_{=0\text{ as shown above}}\mid A\!=\!a\}~~\text{(by independence)}
        \\
        &=\beta_0+\beta_aa+\E[\beta_z(Z)]+\E[\beta_{za}(Z)]a+(\beta_x+\beta_{xa}a)\E[X].
    \end{align*}
    
    Hence weighting by $Q$ results in unbiased ACE estimation in this case,
    $$\E[QY\!\mid\!A\!=\!1]-\E[QY\!\mid\!A\!=\!0]=\beta_a+\E[\beta_{za}(Z)]+\beta_{xa}\E[X].$$
    
    \item[(2)] Now we prove the result for matching on $e(Z,X_{WZA})$. Consider one exposure arm:
    \begingroup
    \allowdisplaybreaks
    \begin{align*}
        &\E\{\E[Y\mid e(Z,X_{WZA}),A=a]\}
        =\E\{\E[Y^{(a)}\mid e(Z,X_{WZA}),A=a]\}\\
        &=\E\{\E[\beta_0+\beta_aa+\beta_z(Z)+\beta_{za}(Z)a+\beta_xX+\beta_{xa}Xa+\epsilon^{(a)}\mid e(Z,X_{WZA}),A=a]\}
        \\
        &=\beta_0+\beta_aa+\E\{\underbrace{\E[\beta_z(Z)\mid e(Z,X_{WZA}),A=a]}_{\E[\beta_z(Z)\mid e(Z,X_{WZA})]}\}+\E\{\underbrace{\E[\beta_{za}(Z)\mid e(Z,X_{WZA},A=a]}_{\E[\beta_{za}(Z)\mid e(Z,X_{WZA})]}\}a+
        \\
        &~~~~(\beta_x+\beta_{xa}a)\E\{\E[X\mid e(Z,X_{WZA}),A=a]\}+\E\{\E[\epsilon^{(a)}\mid e(Z,X_{WZA}),A=a]\}
        \\
        &=\beta_0+\beta_aa+\E[\beta_z(Z)]+\E[\beta_{za}(Z)]a+
        \\
        &~~~~(\beta_x+\beta_{xa}a)\E\{\underbrace{\E[X_{WZA}|e(Z,X_{WZA}),A=a]}_{\E[X_{WZA}\mid e(Z,X_{WZA})]}+\E[X\!-\!X_{WZA}|e(Z,X_{WZA}),A=a]\}+
        \\
        &~~~~\E\{\underbrace{\E[\epsilon^{(a)}\mid Z,X,A=a,W)]}_{\underbrace{\E[\epsilon^{(a)}|Z,X,A\!=\!a]}_{0}~\text{by weak surrogacy}}\}
        \\
        &=\beta_0+\beta_aa+\E[\beta_z(Z)]+\E[\beta_{za}(Z)]a+(\beta_x+\beta_{za}a)\E[X_{WZA}]+
        \\
        &~~~~(\beta_x+\beta_{xa}a)\E\{\underbrace{\E[X\!-\!X_{WZA}|Z,W,A=a]}_{0}\}
        \\
        &=\beta_0+\beta_aa+\E[\beta_z(Z)]+\E[\beta_{za}(Z)]a+(\beta_x+\beta_{za}a)\underbrace{\E\{\E[X|W,Z,A]\}}_{\E[X]}
        \\
        &=\beta_0+\beta_aa+\E[\beta_z(Z)]+\E[\beta_{za}(Z)]a+(\beta_x+\beta_{za}a)\E[X].
    \end{align*}
    \endgroup
    Hence the matching estimator is unbiased for the ACE in this case,
    $$\E\{\E[Y\mid e(Z,X_{WZA}),A=1]-\E[Y\mid e(Z,X_{WZA}),A=0]\}=\beta_a+\E[\beta_{za}(Z)]+\beta_{xa}\E[X].$$
\end{enumerate}
\end{proof}

\newpage


\section*{Supplementary material}

**SUPPLEMENTAL MATERIAL**

for manuscript

**Propensity score analysis when a confounder is a latent variable:**
**Reduction of measurement error bias using the *inclusive* factor score**


This Supplemental Material includes:

A.  Several comments on the use of the proposed method

B.  Implementation steps for incorporating the inclusive factor score (iFS) bias correction method in propensity score (PS) weighting analysis

C.  Mplus inputs for generating the iFS

Mplus inputs for two latent confounders with error correlations can be constructed by combining elements from 1, 2 and 3.

D.  Simulation results for correlated errors

E.  Data example based simulation

F.  Derivation of the $Q_1$ and $H_1$ functions for the logit exposure assignment case with multiple $\boldsymbol{W}$ items that are normally distributed

**A.  Several comments on the use of the proposed method**

A question often asked of any new method is whether, or when, it is necessary. This is a fair question, considering that new methods often require doing things a bit differently from before. The key in deciding whether to use the method is to judge whether the bias matters in the specific case at hand. With a weak confounder and/or small measurement error, the bias may be negligible. If a latent variable is a strong confounder (having strong influence on both exposure assignment and outcome), and measurement error is large, bias correction is important. Large measurement error may result from having few measurement items and/or low item correlations with the latent variable. Also, if the estimated causal effect is borderline significant without using the correction method, one should check if the effect remains when applying the correction. In situations with multiple latent confounders, if (after some reverse-coding if necessary) the latent confounders are positively correlated and have same-sign associations with exposure and same-sign effects on outcome, bias accumulates in the same direction; then bias correction is highly recommended.

To facilitate the incorporation of the iFS bias correction method in PS analysis, we lay out steps for implementation (and sample code) in the sections. Here we highlight two points. First, it is important to establish an appropriate measurement model for the latent variable before applying the iFS bias correction method, as the method presumes that the measurement model is established. While this model may be known for a well researched scale, in other cases it needs to be worked out through careful factor analysis, including measurement invariance testing. Second, when checking balance after PS weighting, for the latent confounder, the balance that needs to be checked is that of the iFS. Balance checks on measurement items or indirect checks on the latent variable (e.g., via testing equality of its mean between two groups) do not capture balance on the iFS.

With regards to computing, we do our analysis in R, and use Mplus to estimate the iFS. This combination is facilitated by the R package MplusAutomation, which makes it simple to call Mplus to estimate the iFS and harvest the iFS to R. While Mplus is a perfect tool for estimating the iFS, it is not required for this purpose; other stand-alone programs and R/Stata/SAS packages that fit SEMs may be explored. The key is to ensure that the program computes FSs conditional on the full model, not just the measurement component of the model. We have not used the gllamm package in Stata for this purpose, but documentation suggests that it would be suitable. And as mentioned earlier, the linear iFS can be computed using software that implements factor models with residual covariances, and does not require fitting SEMs.

A note for Mplus users: Mplus has an option for automatic estimation and saving of PSs when fitting a model with a binary dependent variable. This works well for models with observed covariates. If the PS model has a latent covariate and is fit as a SEM, however, the PS is computed using the measurement-model-based (i.e., conventional) FS (B. O. Muthén, personal communication), and thus suffers from the bias of the cFS proxy method. To obtain the iFS-based PS requires two separate steps: first estimating the iFS, then using the iFS and the observed covariates to estimate the PS.

A special case note: If the application involves a single latent covariate with all continuous measurements and if a logit exposure assignment model is assumed, then one could harvest the relevant parameter estimates (from the same SEM in Fig. 3 in the article) and use the given formulas to compute $X^*$ and the unbiased weighting and matching functions ($Q_1$ and $H_1$), which is theoretically superior to using the iFS and estimating $Q$ and $H$. Since these manual operations are more prone to computing error (than using software-generated FSs), we only recommend this to those who are computing-savvy. Otherwise, it is better to use the iFS. As we have shown, in this case $Q$ and $H$ are numerically very close to $Q_1$ and $H_1$.

**B. Implementation steps for incorporating the iFS bias correction method in PS weighting analysis**

First, consider the simpler case without measurement error (i.e., all confounders are observed and perfectly measured). As a review, in this simple case, PS weighting analysis is conducted as follows:

Step 1: Estimate the PS model using the covariates that we wish to balance, and based on this model, compute the weights

Step 2: In the weighted data, check balance on the covariates. If good balance, proceed. If bad balance, go back to step 1 and modify the PS model.

Step 3: The two options we use in the paper are:

Weighting-only estimator: compute the difference between the weighted mean outcome in the exposed group and the weighted mean outcome in the unexposed group

Weighting-plus estimator: fit an outcome model to the weighted data (regressing the outcome on exposure and covariates), compute for for each individual model-predicted potential outcomes under both exposure and non-exposure, average each potential outcome over the inference group (full sample or the exposed group, demanding on the target causal estimand)

Confidence intervals may be obtained via bootstrapping.

Now consider the case with a latent covariate. To use the iFS method to correct measurement error bias due to the latent covariate, before the steps above, we implement

Step 0.a: Factor analysis to establish the measurement model of the latent covariate, and then measurement invariance testing. We recommend measurement invariance between the exposed and unexposed conditions.

Step 0.b: Estimate the iFS by fitting the relevant SEM and generate the iFS based on this model (see Mplus inputs in section B in this Supplemental Material.)

Next, use the iFS to represent the latent covariate in steps 1-3 of the PS analysis.

## C. Mplus inputs for generating the iFS

Each section includes inputs to generate the logit, probit-ML, probit-WLSMV and linear iFSs

### 1. For a single latent confounder, with uncorrelated measurement errors

This is the simplest of the three cases included.

```
TITLE:     Logit or probit-ML iFS: single latent confounder, uncorrelated measurement errors
DATA:      FILE = "data.dat";
VARIABLE: NAMES    = id z w1 w2 w3 w4 w5 a;   ! customize as needed based on actual data
           IDVARIABLE  = id;
           USEVARIABLE = z w1 w2 w3 w4 w5 a;
           CATEGORICAL = a
                         w1 w2 w3 w4 w5;   ! list Ws here if ordinal, leave out if continuous
ANALYSIS: ESTIMATOR = ML;                  ! either ML or MLR -- does not make a difference in the iFS
           LINK      = PROBIT;             ! use this line if want probit, leave out if want logit
MODEL:     x BY w1* w2 w3 w4 w5; x@1;      ! the latent variable's measurement model
           x ON z;                        ! the model of latent covariate given
           a ON x z;                      ! the exposure assignment model
SAVEDATA: FILE = "factorscore.dat";        ! the iFS will be saved in this file with other variables
           SAVE = FSCORES;                ! (variable names can be found at the end of the .out file)
```

```
TITLE:     Probit-WLSMV iFS: single latent confounder, uncorrelated measurement errors
DATA:      FILE = "data.dat";
VARIABLE: NAMES    = id z w1 w2 w3 w4 w5 a;
           IDVARIABLE  = id;
           USEVARIABLE = z w1 w2 w3 w4 w5 a;
           CATEGORICAL = a
                         w1 w2 w3 w4 w5;   ! list ws here if ordinal, leave out if continuous
ANALYSIS: ESTIMATOR = WLSMV;               ! may omit; Mplus's default estimator w/ categorical vars
MODEL:     x BY w1* w2 w3 w4 w5;  x@1;
           x ON z;
           a ON x z;
SAVEDATA: FILE = "factorscore.dat";
           SAVE = FSCORES;
```

```
TITLE:     Linear iFS by factor model: single latent confounder, uncorrelated measurement errors
DATA:      FILE = "data.dat";
VARIABLE: NAMES    = id z w1 w2 w3 w4 w5 a;
           IDVARIABLE  = id;
           USEVARIABLE = z w1 w2 w3 w4 w5 a;
MODEL:     x BY w1* w2 w3 w4 w5 z a;  x@1;
           z WITH a;
SAVEDATA: FILE = "factorscore.dat";
           SAVE = FSCORES;
```

If using complex sample data,
- in the ANALYSIS statement, add TYPE = COMPLEX and, if want the logit or probit-ML factor score versions, change ESTIMATOR from ML to MLR; and
- in the VARIABLE statement, add CLUSTER, STRATIFICATION and WEIGHT variables to describe the sampling design as appropriate.

For how to deal with other types of W variables (count, unordered categorical), see Mplus User's Guide.

## 2. *For a single latent confounder, with correlated measurement errors*

These inputs are modified based on those in 1. Here we assume dependence between the error terms in the first two measurement items.

Note: when the nuisance factor code is used, two FSs will be generated, one for X and one for the nuisance factor S. Only the one for X is needed for the PS analysis.

```
TITLE:     Logit or probit-ML iFS: single latent confounder, correlated measurement errors
DATA:      FILE = "data.dat";
VARIABLE:  NAMES      = id z w1 w2 w3 w4 w5 a;
           IDVARIABLE = id;
           USEVARIABLE = z w1 w2 w3 w4 w5 a;
           CATEGORICAL = a
                         w1 w2 w3 w4 w5;  ! list ws here if ordinal, leave out if continuous
ANALYSIS:  ESTIMATOR = ML;               ! either ML or MLR -- does not make a difference in the iFS
           LINK      = PROBIT;           ! use this line if want probit, leave out if want logit
MODEL:     x BY w1* w2 w4 w5; x@1;
           x ON z;
           a ON x z;
           ! below are 2 options for conditional dependence of W1, W2 given X, Z. pick one.
           w1 WITH w2;                   ! option 1 (ws continuous only): error covariance
           s BY w1* (1);  s BY w2 (1);  s@1;  ! option 2 (ws ordinal ok): nuisance factor S that is
           s WITH x@0 z@0;  a ON s@0;    !                  uncorrelated with X,Z,A
SAVEDATA:  FILE = "factorscore.dat";
           SAVE = FSCORES;
```

```
TITLE:     Probit-WLSMV iFS: single latent confounder, correlated measurement errors
DATA:      FILE = "data.dat";
VARIABLE:  NAMES      = id z w1 w2 w3 w4 w5 a;
           IDVARIABLE = id;
           USEVARIABLE = z w1 w2 w3 w4 w5 a;
           CATEGORICAL = a
                         w1 w2 w3 w4 w5;  ! list ws here if ordinal, leave out if continuous
ANALYSIS:  ESTIMATOR = WLSMV;            ! may omit; Mplus's default estimator for categorical vars
MODEL:     x BY w1* w2 w4 w5;  x@1;
           x ON z;
           a ON x z;
           ! below are 2 options for conditional dependence of W1, W2 given X, Z. pick one.
           w1 WITH w2;                   ! option 1 (continuous/ordinal Ws): error covariance
           s BY w1* (1);  s BY w2 (1);  s@1;  ! option 2 (ordinal Ws): nuisance factor S that is
           s WITH x@0 z@0;  a ON s@0;    !                  uncorrelated with X,Z,A
SAVEDATA:  FILE = "factorscore.dat";
           SAVE = FSCORES;
```

```
TITLE:     Linear iFS by factor model: single latent confounder, correlated measurement errors
DATA:      FILE = "data.dat";
VARIABLE:  NAMES      = id z w1 w2 w3 w4 w5 a;
           IDVARIABLE = id;
           USEVARIABLE = z w1 w2 w3 w4 w5 a;
MODEL:     x BY w1* w2 w3 w4 w5 z a;  x@1;
           z WITH a;
           w1 WITH w2;  ! error covariance is the only option, as this model treats ws as continuous
SAVEDATA:  FILE = "factorscore.dat";
           SAVE = FSCORES;
```

### 3.  *For two latent confounders, with uncorrelated measurement errors*

These inputs are modified based on those in 1. Here we have two latent variables, X1 and X2.

```
TITLE:      Logit or probit-ML iFS: two latent confounders, uncorrelated measurement errors
DATA:       FILE = "data.dat";
VARIABLE:   NAMES       = id z w1 w2 w3 w4 w5 w6 w7 w8 w9 w10 a;
            IDVARIABLE  = id;
            USEVARIABLE = z w1 w2 w3 w4 w5 w6 w7 w8 w9 w10 a;
            CATEGORICAL = a
                          w1 w2 w3 w4 w5 w6 w7 w8 w9 w10;  ! list if ordinal, leave out if continuous
ANALYSIS:   ESTIMATOR = ML;                   ! either ML or MLR – does not make a difference in the iFS
            LINK      = PROBIT;               ! use this line if want probit, leave out if want logit
MODEL:      x1 BY w1* w2 w3 w4 w5;    x1@1;
            x2 BY w6* w7 w8 w9 w10;   x2@1;
            x1 WITH x2;
            x1 x2 ON z;
            a ON x1 x2 z;
SAVEDATA:   FILE = "factorscore.dat";
            SAVE = FSCORES;


TITLE:      Probit-WLSMV iFS: two latent confounders, uncorrelated measurement errors
DATA:       FILE = "data.dat";
VARIABLE:   NAMES       = id z w1 w2 w3 w4 w5 w6 w7 w8 w9 w10 a;
            IDVARIABLE  = id;
            USEVARIABLE = z w1 w2 w3 w4 w5 w6 w7 w8 w9 w10 a;
            CATEGORICAL = a
                          w1 w2 w3 w4 w5 w6 w7 w8 w9 w10;  ! list if ordinal, leave out if continuous
ANALYSIS:   ESTIMATOR = WLSMV;               ! may omit; Mplus's default estimator for categorical vars
MODEL:      x1 BY w1* w2 w3 w4 w5;    x1@1;
            x2 BY w6* w7 w8 w9 w10;   x2@1;
            x1 WITH x2;
            x1 x2 ON z;
            a ON x1 x2 z;
SAVEDATA:   FILE = "factorscore.dat";
            SAVE = FSCORES;


TITLE:      Linear iFS by factor model: two latent confounders, uncorrelated measurement errors
DATA:       FILE = "data.dat";
VARIABLE:   NAMES       = id z w1 w2 w3 w4 w5 w6 w7 w8 w9 w10 a;
            IDVARIABLE  = id;
            USEVARIABLE = z w1 w2 w3 w4 w5 w6 w7 w8 w9 w10 a;
MODEL:      x1 BY w1* w2 w3 w4 w5 z a;    x1@1;
            x2 BY w6* w7 w8 w9 w10 z a;   x2@1;
            z WITH a;
            x1 WITH x2;
SAVEDATA:   FILE = "factorscore.dat";
            SAVE = FSCORES;
```

## D. Simulation results for correlated errors

Using several scenarios as examples (the range of simulation scenarios is described in the paper), the table below shows that the iFS that accounts for dependence among measurement errors performs well, as expected. The FS method that is otherwise inclusive but incorrectly assumes that the error terms of $W_1$ and $W_2$ are independent is biased in estimating the ACE, but is less biased than using the measurement items.

Table C1: The iFS that correctly accommodates measurement error dependence, compared to one that assumes independence, and compared to using $\overline{W}$, and to the true $X$

| Scenario | $W$ type | Mean estimate when using | | | |
|---|---|---|---|---|---|
| | | $X$ | $\overline{W}$ | incorrect iFS (assuming error independence) | correct iFS (accommodating error dependence) |
| base | continuous | 0.004 | 0.152 | 0.049 | 0.005 |
| | ordinal | 0.004 | 0.173 | 0.045 | 0.004 |
| double effect of $X$ on $A$ | continuous | 0.004 | 0.299 | 0.093 | 0.005 |
| | ordinal | 0.004 | 0.341 | 0.084 | 0.003 |
| double effect of $X$ on $Y$ | continuous | 0.006 | 0.297 | 0.090 | 0.005 |
| | ordinal | 0.006 | 0.337 | 0.087 | 0.008 |

## E. Data example based simulation

For a simulation study that is as relevant as possible to the real data example, we retain the same sample and all the same observed covariates with their observed values, and generate the remaining variables using (logit-SEM-estimated) parameters based on the real data.

We generate the two latent covariates aa (which stands for academic achievement) and vv (violence) based on a linear model on the observed covariates (White, Black, American-Indian, Asian, Other race, Hispanic ethnicity, age, parent education less than high school, high school, business/vocational training, some college, college degree, parent being single, widowed, divorced and separated). The mean components of aa and vv depends on these observed covariates via the sets of coefficients (.167, .005, .085, .992, .161, .229, .022, -.369, -.365, -.432, -.108, .147, -.532, -.304, -.280) and (-.286, -.143, .361, -.286, -.653, .338, -.059, .044, .208, .467, .041, .097, -.131, .139, .250), respectively. The error component of these two variables is multivariate normal, with zero mean, unit variance, and covariance .234.

To generate the measurement items for these two latent covariates, we first generate their underlying continuous versions based on the (unstandardized) factor loadings of these variables and standard logistic errors, and then convert these into the desired categorical versions using thresholds.

|  | Loading on aa | Thresholds |  | Loading on vv | Thresholds |
|---|---|---|---|---|---|
| Math | 1.919 | -1.996, .478, 3.305 | Fight | 1.936 | -2.207, .149, 1.276 |
| English | .824 | -.1.000, .467, 1.926 | Hurt in fights | 2.169 | -1.099, 1.553, 2.511 |
| Social sciences | .953 | -1.425, .255, 1.964 | Weapon use | 1.456 | 1.536, 2.856, 3.361 |
| Natural sciences | 1.267 | -1.344, .456, 2.288 | Group fight | 1.994 | -.760, 1.594, 2.745 |

The exposure variable is generated as a random binary variable with probability based on a logit model of the observed covariates (with coefficients .589, .456, .640, .447, -.161, .028, -.066, .009, .290, -.105, -.757, 1.442, -.353, -.535, .229) and the true values of aa and vv (coefficients -.607 and .306), with threshold -.169.

Analysis models will use the logit link for the exposure, which is correctly specified. Analysis models with the latent variables' measurement components use the logit link for the measurement items, which is also correctly specified.

We generate outcomes using three different models. The first model is a logit model using the SEM-estimated parameters: coefficients .488, .168, .212, .933, -1.247, .360, .049, -.119, .303, .130, .231, .517, .358, -.512, .180, .530 for the observed covariates and -.266 and .287 for the true values of aa and vv, and threshold -1.964. This provides for the weighting-plus estimator's use of the logit-linear outcome model to be the correct model. The second outcome model we use modifies the model above by shifting the linear predictor by $-e^{aa}/10$. In this case the weighting-plus estimator's logit-linear outcome model is misspecified. The third outcome model uses a probit instead of logit model; it borrows the parameters of the first model but rescales them by a factor of 1/1.7. In this case the weighting-plus estimator's logit-linear outcome model is also misspecified.

We compare results from analyses (1) with no measurement error bias correction, i.e., using the average of the aa items and the average of the vv items (mean scores) to represent the latent covariates; (2) correcting for aa only or vv only, i.e., applying the iFS method to one latent variable, assuming the other one is observed (and using its mean score); (3) correcting for both.

With the estimand being the ACEE, the inference population (the exposed group) changes from one simulated dataset to another. There is thus no one true ACEE value. Each dataset has one ACEE value, and the closest we can get to that value is the value estimated using the actual values of the latent variables aa and vv, i.e., in the absence of measurement error. With each dataset, for each estimator, we track its deviation from the true ACEE as well as its deviation from the value that would be obtained in the absence of measurement error. We present below, for each outcome scenario and each estimator, the arithmetic mean and quadratic mean of these deviations over the 2000 simulated datasets; these two metrics are analogous to bias and RMSE for a single estimand.

| | Deviations from the true ACEE | | | | Deviations from the estimate using aa, vv | | | |
| | weighting-only estimator | | weighting-plus estimator | | weighting-only estimator | | weighting-plus estimator | |
| | arithmetic mean | quadratic mean | arithmetic mean | quadratic mean | arithmetic mean | quadratic mean | arithmetic mean | quadratic mean |
|---|---|---|---|---|---|---|---|---|
| **Outcome 1: logit-linear** | | | | | | | | |
| no correction | 2.24 | 6.42 | 2.24 | 6.17 | 2.15 | 3.12 | 2.19 | 2.94 |
| vv corrected | 1.31 | 6.36 | 1.36 | 6.03 | 1.22 | 2.75 | 1.31 | 2.46 |
| aa corrected | 0.79 | 6.78 | 0.61 | 6.30 | 0.70 | 3.02 | 0.56 | 2.57 |
| both corrected | 0.05 | 6.88 | -0.07 | 6.37 | -0.03 | 3.06 | -0.12 | 2.59 |
| true aa, vv | 0.09 | 6.26 | 0.05 | 5.91 | | | | |
| **Outcome 2: logit-nonlinear** | | | | | | | | |
| no correction | 2.78 | 6.36 | 2.78 | 6.11 | 2.80 | 3.61 | 2.85 | 3.46 |
| vv corrected | 1.86 | 6.24 | 1.93 | 5.93 | 1.88 | 3.11 | 1.99 | 2.92 |
| aa corrected | 0.72 | 6.50 | 0.46 | 5.98 | 0.73 | 3.03 | 0.53 | 2.61 |
| both corrected | 0.03 | 6.62 | -0.16 | 6.07 | 0.05 | 3.04 | -0.09 | 2.61 |
| true aa, vv | -0.02 | 5.84 | -0.07 | 5.48 | | | | |
| **Outcome 3: probit-linear** | | | | | | | | |
| no correction | 2.13 | 6.37 | 2.11 | 6.10 | 2.09 | 3.10 | 2.15 | 2.90 |
| vv corrected | 1.26 | 6.34 | 1.30 | 6.05 | 1.22 | 2.76 | 1.34 | 2.46 |
| aa corrected | 0.73 | 6.72 | 0.52 | 6.24 | 0.69 | 3.03 | 0.57 | 2.55 |
| both corrected | 0.06 | 6.84 | -0.09 | 6.31 | 0.02 | 3.05 | -0.05 | 2.54 |
| true aa, vv | 0.04 | 6.18 | -0.05 | 5.80 | | | | |

These results show that for both the weighting-only and weighting-plus estimators there is a decrease in the estimated ACE we move from no measurement error bias correction, to correction for one latent covariate, to correction for both covariates (using the iFS method), and that correction for both covariates obtains ACE estimates that are closest to estimates that would be obtained if the true values of the latent variable were used.

## F. Derivation of the $Q_1$ and $H_1$ functions for the logit exposure assignment case with multiple $\boldsymbol{W}$ items that are normally distributed

First, let's start with the simple single measurement case where $W = X + U$, $U \sim \mathrm{N}(0, \sigma^2)$, and $\mathrm{P}(A = 1 \mid Z, X) = \{1 + \exp[-(\beta_0 + \beta_x X + \beta_z Z)]\}^{-1}$ in McCaffrey et al. (2013, pp. 674). For units with $A = 1$ and units with $A = 0$, the inverse probability weights are $1 + \exp[-(\beta_0 + \beta_x X + \beta_z Z)]$ and $1 + \exp(\beta_0 + \beta_x X + \beta_z Z)$, respectively. Combining these, the true weight function is

$$Q_0 = 1 + \exp[(1 - 2A)(\beta_0 + \beta_x X + \beta_z Z)].$$

Consider the exponential. Leveraging the relationship $X = W - U$, replace $X$ with $W$ to obtain

$$\begin{aligned}
\exp[(1 - 2A)(\beta_0 + \beta_x W + \beta_z Z)] &= \exp[(1 - 2A)(\beta_0 + \beta_x X + \beta_x U + \beta_z Z)] \\
&= \exp[(1 - 2A)(\beta_0 + \beta_x X + \beta_z Z)] \cdot \exp[(1 - 2A)\beta_x U].
\end{aligned}$$

Taking expectations conditional on $(Z, X, A)$ obtains

$$\begin{aligned}
\mathrm{E}\{\exp[(1 - 2A)&(\beta_0 + \beta_x W + \beta_z Z)] \mid Z, X, A\} \\
&= \exp[(1 - 2A)(\beta_0 + \beta_x X + \beta_z Z)] \cdot \mathrm{E}\{\exp[(1 - 2A)\beta_x U] \mid Z, X, A\} \\
&= \exp[(1 - 2A)(\beta_0 + \beta_x X + \beta_z Z)] \cdot \mathrm{E}\{\exp[(1 - 2A)\beta_x U]\} \\
&= \exp[(1 - 2A)(\beta_0 + \beta_x X + \beta_z Z)] \cdot \exp(\beta_x^2 \sigma^2 / 2),
\end{aligned}$$

where the last equality is due to that fact that $\exp[(1 - 2A)\beta_x U] \sim \text{log-normal}(0, \beta_x^2 \sigma^2)$. Let

$$Q_1 = 1 + \exp[(1 - 2A)(\beta_0 + \beta_x W + \beta_z Z)] \cdot \exp(-\beta_x^2 \sigma^2 / 2)$$

(the weighting function defined in McCaffrey et al., 2013). Then $\mathrm{E}[Q_1 \mid Z, X, A] = Q_0$. That is, $Q_1$ is an unbiased weighting function. It can be abbreviated to

$$Q_1 = 1 + \exp[(1 - 2A)(\beta_0 + \beta_x X^* + \beta_z Z)], \text{ where } X^* = W + (2A - 1)\beta_x \sigma^2 / 2.$$

Let's now switch to the case with multiple measurement items. With the same logit exposure assignment model, $Q_0$ is the same as above. Now $\boldsymbol{W} = \boldsymbol{\lambda}_0 + \boldsymbol{\lambda}_x X + \boldsymbol{\lambda}_z Z + \boldsymbol{U}$ where $\boldsymbol{U} \sim \mathrm{N}(\boldsymbol{0}, \boldsymbol{\Sigma})$. ($\boldsymbol{\lambda}_z$ is only needed if some $\boldsymbol{W}$ items are dependent on $Z$ given $X$.) Now we leverage the relationship

$$\begin{aligned}
X &= (\boldsymbol{\lambda}_x' \boldsymbol{\Sigma}^{-1} \boldsymbol{\lambda}_x)^{-1} \boldsymbol{\lambda}_x' \boldsymbol{\Sigma}^{-1} (\boldsymbol{W} - \boldsymbol{\lambda}_0 - \boldsymbol{\lambda}_z Z - \boldsymbol{U}) \\
&= (\boldsymbol{\lambda}_x' \boldsymbol{\Sigma}^{-1} \boldsymbol{\lambda}_x)^{-1} \boldsymbol{\lambda}_x' \boldsymbol{\Sigma}^{-1} (\boldsymbol{W} - \boldsymbol{\lambda}_0 - \boldsymbol{\lambda}_z Z) - (\boldsymbol{\lambda}_x' \boldsymbol{\Sigma}^{-1} \boldsymbol{\lambda}_x)^{-1} \boldsymbol{\lambda}_x' \boldsymbol{\Sigma}^{-1} \boldsymbol{U}.
\end{aligned}$$

Denote the first of the two terms above by $X_{\mathrm{MLE}}$; this label is appropriate because this term (like $W$ in the single measurement case) is the MLE of $X$ based on the measurement model

(the $\boldsymbol{W}|X, Z$ model) – if we treat model parameters $(\boldsymbol{\lambda}_0, \boldsymbol{\lambda}_x, \boldsymbol{\lambda}_z, \boldsymbol{\Sigma})$ as known and treat the $X$ value of each unit as an unknown parameter. Replace $X$ in the exponential in the $Q_0$ formula with $X_{\text{MLE}}$ to obtain

$$\begin{aligned}
\exp[(1-2A)(\beta_0 + \beta_x X_{\text{MLE}} + \beta_z Z)] = {} & \\
= \exp[(1-2A)(\beta + \beta_x X + \beta_z Z)] & \cdot \exp[(1-2A)\beta_x(\boldsymbol{\lambda}_x'\boldsymbol{\Sigma}^{-1}\boldsymbol{\lambda}_x)^{-1}\boldsymbol{\lambda}_x'\boldsymbol{\Sigma}^{-1}\boldsymbol{U}].
\end{aligned}$$

Taking expectations conditional on $(Z, X, A)$ obtains

$$\begin{aligned}
\text{E}\{\exp[(1-2A)(\beta_0 + \beta_x X_{\text{MLE}} + \beta_z Z)] \mid Z, X, A\} = {} & \\
= \exp[(1-2A)(\beta + \beta_x X + \beta_z Z)] & \cdot \text{E}\{\exp[(1-2A)\beta_x(\boldsymbol{\lambda}_x'\boldsymbol{\Sigma}^{-1}\boldsymbol{\lambda}_x)^{-1}\boldsymbol{\lambda}_x'\boldsymbol{\Sigma}^{-1}\boldsymbol{U}]\} \\
= \exp[(1-2A)(\beta + \beta_x X + \beta_z Z)] & \cdot \exp[\beta_x^2(\boldsymbol{\lambda}_x'\boldsymbol{\Sigma}^{-1}\boldsymbol{\lambda}_x)^{-1}/2],
\end{aligned}$$

since $\exp[(1-2A)\beta_x(\boldsymbol{\lambda}_x'\boldsymbol{\Sigma}^{-1}\boldsymbol{\lambda}_x)^{-1}\boldsymbol{\lambda}_x'\boldsymbol{\Sigma}^{-1}\boldsymbol{U}] \sim \text{log-normal}(0, \beta_x^2(\boldsymbol{\lambda}_x'\boldsymbol{\Sigma}^{-1}\boldsymbol{\lambda}_x)^{-1})$. Define

$$Q_1 = 1 + \exp[(1-2A)(\beta_0 + \beta_x X_{\text{MLE}} + \beta_z Z)] \cdot \exp[-\beta_x^2(\boldsymbol{\lambda}_x'\boldsymbol{\Sigma}^{-1}\boldsymbol{\lambda}_x)^{-1}/2].$$

Then $\text{E}[Q_1 \mid Z, X, A] = Q_0$. That is, $Q_1$ is the unbiased weighting function. This can also be abbreviated,

$$Q_1 = 1 + \exp[(1-2A)(\beta_0 + \beta_x X^* + \beta_z Z)], \text{ where } X^* = X_{\text{MLE}} + (2A-1)\beta_x(\boldsymbol{\lambda}_z'\boldsymbol{\Sigma}^{-1}\boldsymbol{\lambda}_x)^{-1}/2.$$

Based on Carroll et al. (2006, eq. 7.12), $e(Z, X^*) = \{1 + \exp[-(\beta_0 + \beta_x X^* + \beta_z Z)]\}^{-1}$, so $Q_1$ is an inverse probability weighting function. Since $Q_1$ is an unbiased weighting function, it follows (by the connection between unbiased weighting and matching functions shown in Lockwood & McCaffrey, 2016) that $H_1 = (Z, X^*)$ and $H_1 = e(Z, X^*)$ are unbiased matching functions.



\end{document}